\documentclass[12pt,preprint]{aastex}
\shorttitle{supporting system}
\shortauthors{Zhang et al.}

\begin{document}

\title{Astronomical Observing Conditions at Xinglong Observatory from 2007 to 2014}

\author{JI-CHENG ZHANG$^{1,2}$, LIANG GE$^{2}$, XIAO-MENG LU$^{2}$, ZI-HUANG CAO$^{2}$, XU CHEN$^{1}$, YONG-NA MAO$^{2}$, XIAO-JUN JIANG$^{2}$}
\affil {1.Shandong Provincial Key Laboratory of Optical Astronomy and Solar-Terrestrial Environment, Institute of Space Sciences, School of Space Science and Physics,Shandong University, Weihai, 264209, China, jczhang@bao.ac.cn}
\affil {2.Key Laboratory of Optical Astronomy, National Astronomical Observatories, Chinese Academy of Sciences, Beijing 100012, China}

\begin{abstract}
  Xinglong observatory of National Astronomical Observatories, Chinese Academy of Sciences (NAOC) is one of the major optical observatories in China, which hosts nine optical telescopes including the Large Sky Area Multi-Object Fiber Spectroscopic Telescope (LAMOST) and the 2.16-m reflector. Scientific research from these telescopes are focused on stars, galaxies and exo-planets using multi-color photometry and spectroscopic observations. Therefore, it is important to provide the observing conditions of the site, in detail, to the astronomers for an efficient use of these facilities. In this paper, we present the characterization of observing conditions at Xinglong observatory based on the monitoring of meteorology, seeing and sky brightness during the period from 2007 to 2014. Meteorological data were collected from a commercial Automatic Weather Station (AWS), calibrated by China Meteorological Administration. Mean and median wind speed are almost constant during the period analysed and ranged from 1.0 m s$^{-1}$ to 3.5 m s$^{-1}$. However, high wind speed ($\geq$ 15 m s$^{-1}$) interrupts observations, mainly, during winter and spring. Statistical analysis of air temperature showed the temperature difference between daytime and nighttime, which can be solved by opening the ventilation device and the slit of the dome at least one hour before observations. Analysis resulted in average percentage of photometric nights and spectroscopic nights are 32\% and 63\% per year, respectively. The distribution of photometric nights and spectroscopic nights have significant seasonal tendency, worsen in summer due to clouds, dust, and high humidity. Seeing measurements were obtained using the Differential Image Motion Monitor (DIMM). Mean and median value of seeing over one year are around 1.9$''$ and 1.7$''$, respectively. 80\% of nights with seeing values are below 2.6$''$ whereas the distribution peaks around 1.8$''$. The measurements of sky brightness are acquired from Sky Quality Meter (SQM) and photometric observations. Analysis shows that sky brightness at zenith is around 21.1 mag arcsec$^{-2}$ and becomes brighter with large zenith angle. Sky brightness increases due to the light pollution of surrounding cities, e.g. Beijing, Tangshan and Chengde. Significant influence towards the direction of Beijing, at an altitude of $30^{\circ}$, can increase the sky brightness up to 20.0 mag arcsec$^{-2}$. Sky brightness reduces after midnight, mainly because of the influence of city lights and the artificial acts. The above results suggest that Xinglong observatory is still a good site for astronomical observations. Our analysis of the observing conditions at Xinglong observatory can be used as a reference to the observers on targets selection, observing strategy, and telescope operation.
\end{abstract}

\keywords{Meteorology, Seeing, Sky Brightness}

\section{INTRODUCTION}

\begin{figure*}[h]
\centering
\includegraphics[angle=0,scale=0.5]{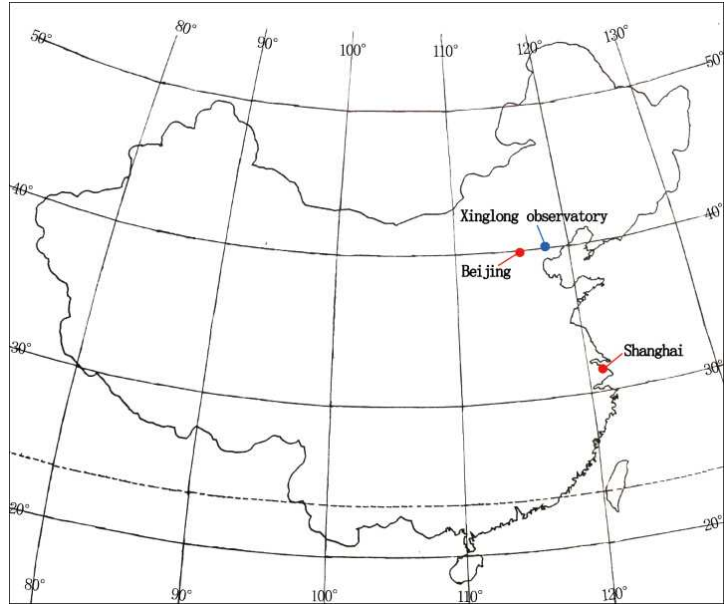}
\hfill \caption{Location of Xinglong observatory is shown in the map. Beijing, major city near to observatory, also located for reference. See the electronic edition of the PASP for a color version of this figure.}
\label{fig1}
\end{figure*}

Xinglong observatory (117$^{\circ}$34$'$39$''$ East, 40$^{\circ}$23$'$26$''$ North)(see Figure~\ref{fig1}) of National Astronomical Observatories, Chinese Academy of Sciences (NAOC) is located 120km northeast to Beijing, at an altitude of about 900 meter on south of the main peak of Yanshan Mountains. It belongs to the typical monsoon climate with northwest monsoon in winter and southeast monsoon in summer. Temperature range from -20$^\circ$C in winter to 30$^\circ$C in summer. It contains nine optical telescopes, including the Large Sky Area Multi-Object Fiber Spectroscopic Telescope (LAMOST) \citep{2012RAA....12.1197C} and the 2.16-meter reflector. As one of major optical observatories in China, it is important to give its observing conditions in detail as a reference for astronomers. Previous researches and references on the observing conditions at Xinglong observatory have been mainly mentioned at \cite{2012RAA....12..772Y}, these measurements could not reflect the seeing value from DIMM with monthly variations and the sky brightness variations with different altitude and azimuth systematically. In this paper we will focus on three important characterizations of an optical observatory, the meteorology, the seeing and the sky brightness.

For ground-based astronomy, fraction of clear nights, seeing, and sky brightness are important parameters to evaluate the quality of an observing site \citep{2003A&A...400.1183P}. Many astronomical observatories in the world have given detailed studies of these parameters. \cite{1989BASI...17...83S} monitored the sky and atmospheric conditions at Leh, including the photometric and spectroscopic hours, meteorology, seeing conditions and extinction coefficients. \cite{2004PASP..116..762T} presented the sky brightness (UBVR bands) and the seeing measurements of the Mount Graham International Observatory, which contained the Vatican Advanced Technology Telescope (VATT), the Heinrich Hertz Submillimeter Telescope, and the Large Binocular Telescope (LBT). Thirty Meter Telescope (TMT) site-testing group suggested a detailed list of atmospheric parameters which should be measured, including weather-related characteristics (e.g. fraction of cloud cover and photometric conditions, air temperature and ground-level humidity.), turbulence-related characteristics (e.g. overall seeing, isoplanatic angle.) and Other characteristics (e.g. dust, sky brightness, atmospheric transparency.) \citep{2009PASP..121..384S}. The site-testing group for the future European Extremely Large Telescope (E-ELT) also selected a list of key parameters which should be measured ideally, including turbulence-related features, meteorological conditions (e.g. wind speed and direction, relative humidity, air temperature,atmospheric extinction,cloud cover and dust.) and other parameters (e.g. sky brightness, sodium layer.) \citep{2011PASP..123.1334V}. Astronomical site evaluation in the visible and radio range of an IAU technical workshop presented a systematic introduction on atmospheric turbulence, site surveys and forecasting, etc. A detailed description was given by \cite{2002ASPC..266.....V}.

Definition of clear nights is judged by the quantitative proportion of cloud coverage or the time duration of no clouds through the whole night, criterion of the percentage of cloud coverage at the world observatories are different from each other. \cite{2000A&AS..145..293E} defined $''$clear time$''$ at the Maidanak Observatory as cloud coverage smaller than 25\%. \cite{1989BASI...17...83S} labelled nights with at least six uninterrupted hours of photometric sky conditions as $''$clear nights$''$. Cloud coverage is a crucial parameter for determining the useful time in an astronomical observatory. The clear nights are judged mainly on visual basis during earlier days, however, this method is only suitable for the sites where there are assistant observers but not for identifying future candidate sites. A new methodology which can quantify the fraction of clear nights by the satellite images for La Palma and Mt Graham in \cite{2011MNRAS.411.1271C}, presented the ground-based data as a reference simultaneously. This measurement was demonstrated for monitoring the methodology at astronomical sites successfully in \cite{2002ASPC..266..310E}. Using satellites for cloud coverage analysis has also been applied and demonstrated in other works (Erasmus and Maartens, Operational forecasts of cirrus cloud cover and water vapour above Paranal and la Silla observatories, Purchase Order 52538/VPS/97/9480/HWE, 1999; Erasmus and Van Rooyen, A Satellite Survey of Cloud Cover and Water Vapour in Morocco and Southern Spain and a Verification Using La Palma Ground-base Observations, Purchase Order 73526/TSD/04/6179/GWI/LET, 2006; Varela et al., On the Use of Remotely Sensed Data for Astronomical Site Characterization.).

Seeing is vital to high precision imaging and high spatial resolution photometry for ground-based astronomy. Astronomical seeing refers to the optical inhomogeneities in the earth's atmosphere \citep{1974ApJ...189..587Y}, which is usually measured by using Differential Image Motion Monitor (DIMM), this is now commonly used to measure seeing, \cite{1990A&A...227..294S} described the DIMM theory for the first time, DIMM seeing was measured by the variance of the differential motion of two sub-aperture images of a star. A detailed description of DIMM with design, data processing and calibration was given by \cite{1995PASP..107..265V}. Main advantage on seeing calculation by DIMM is that the tracking errors are subtracted automatically, however the defocus (www.alcor-system.com/us/DimmSoftware/defoc.pdf), spot saturation \citep{2004SPIE.5382..656V} and the optical quality of telescope could also introduce important errors in seeing calculation by DIMM. During the study on theory of DIMM thoroughly, more accurate coefficients (e.g. CCD noise and exposure time) have been taken into consideration by \cite{2002PASP..114.1156T}. This theory and instrument has been widely used in astronomical sites all over the world (e.g. \cite{1999AstL...25..122I}, \cite{2005MNRAS.362..455Z}, \cite{2010PASP..122..731F}).

Sky brightness is crucial for optical observatories as it restricts not only the limitation of telescope, but also the sample frequency for given targets and the signal-to-noise ratio(SNR) for same telescope with certain exposure time. In order to protect the astronomical quality and prevent light pollution, the sky law for the protection on the island of La Palma and the area of Tenerife has been passed by the Spanish government since 1988. Many observatories in the world have given the studies of sky brightness. \cite{2000PASP..112..566M} measured the sky brightness over the Mount Hopkins and Kitt Peak with spectrophotometry, then convected to broadband measurements for the zenith, they also gave the variations with different directions corresponding to the surrounding cities in a time span of a decade. \cite{1995A&AS..112...99L} and \cite{2007PASP..119.1186S} presented the measurements of sky brightness with optical spectrophotometric data and photometric-calibrated data at Calar Alto observatory, which has the largest telescope in continental Europe. \cite{2003A&A...400.1183P} and \cite{2008A&A...481..575P} provided the sky brightness with the influence of the time span and the solar activities at Cerro Paranal, which is the site of the Very Large Telescopes (VLT). \cite{2014PASP..126.1068A} evaluated the sky brightness of two candidate Argentinian observation sites for Cherenkov Telescope Array with Sky Quality Meters (SQM) and the spectrometer for night aerosol detection.

This paper consists of five sections. Section 1 gives a brief introduction. We describe the meteorology of Xinglong observatory in Section 2. Section 3 presents the seeing data, which is measured by DIMM. In Section 4, we describe the sky brightness of Xinglong observatory and its variations. Section 5 presents the summary and conclusions.

\section{METEOROLOGICAL ANALYSIS AT XINGLONG OBSERVATORY}

We present a statistical analysis on air temperature, wind speed, wind direction, relative humidity, and fraction of photometric nights and spectroscopic nights in this section. All the meteorological data were collected from the Vaisala HydroMet $^{TM}$ System MAWS110, this is a commercial Automatic Weather Stations (AWS) which includes sensors of air-temperature, humidity, pressure, wind speed and direction. These sensors were installed on towers 10m above the ground and calibrated by the China Meteorological Administration. Air temperature sensor gives an accuracy of $\pm$0.2$^\circ$C. Accuracy of wind speed is $\pm$0.2 m s$^{-1}$ with the range from 0.4 m s$^{-1}$ to 60 m s$^{-1}$, accuracy of the wind direction is $\pm$2$^\circ$. Relative humidity gives an accuracy of $\pm$2\% below 90\%, and $\pm$3\% above 90\%. This system is especially designed for unattended operations requiring high reliability and accuracy at sites with the mains power and with battery back-up. The MAWS110 uses a field proven and high accuracy data logger and advanced software, all the meteorological data are sampled every two minutes by software from the AWS.

\subsection{Air Temperature}

Air temperature is an important parameter in operating the telescope and its detector, e.g. charge coupled device (CCD). It is known that CCD generate thermal electrons (also known as dark current), related to its operating temperature. In order to set the temperature of CCD, cooling is essential. Type of CCD, cooled with thermoelectric coolers, is influenced by air temperature.

\begin{figure*}[htbp]
\centering
\includegraphics[angle=0,scale=0.35]{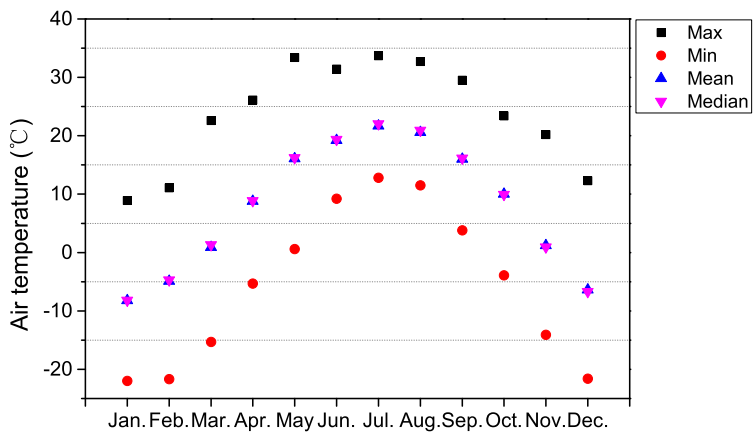}
\hfill \caption{Monthly statistics of the air temperature for all years together from 2007 to 2014 are shown. Black squares, red circles, blue triangles, and carmine inverted triangles indicate the maximum, minimum, mean ,and median values, respectively. See the electronic edition of the PASP for a color version of this figure.}
\label{fig2}
\end{figure*}

We have analysed the distribution of air temperature based on monthly basis and calculated the maximum, minimum, mean and median values that are shown in Figure~\ref{fig2}. The air temperature occupies the range, 9.2$^\circ$C to 33.7$^\circ$C, and show higher in summer compared to other seasons (In this paper, we defined local seasons at Xinglong observatory by grouping calendar months as follows. spring: March, April and May; summer: June, July and August; autumn: September, October and November; winter: December, January and February.). Monthly differences between maximum value and minimum value have the same tendency, but mean value and median value of air temperature are almost same. The measurements based on 24hours, daytime and nighttime analysis along with number of data (Ndata) are summarized in Table 1,2 and 3, respectively. Statistical analysis of air temperature showed the temperature difference between daytime and nighttime, which can be solved by opening the ventilation device and dome slit at least one hour before the start of observations.

\begin{figure*}[htbp]
\centering
\includegraphics[angle=0,scale=0.25]{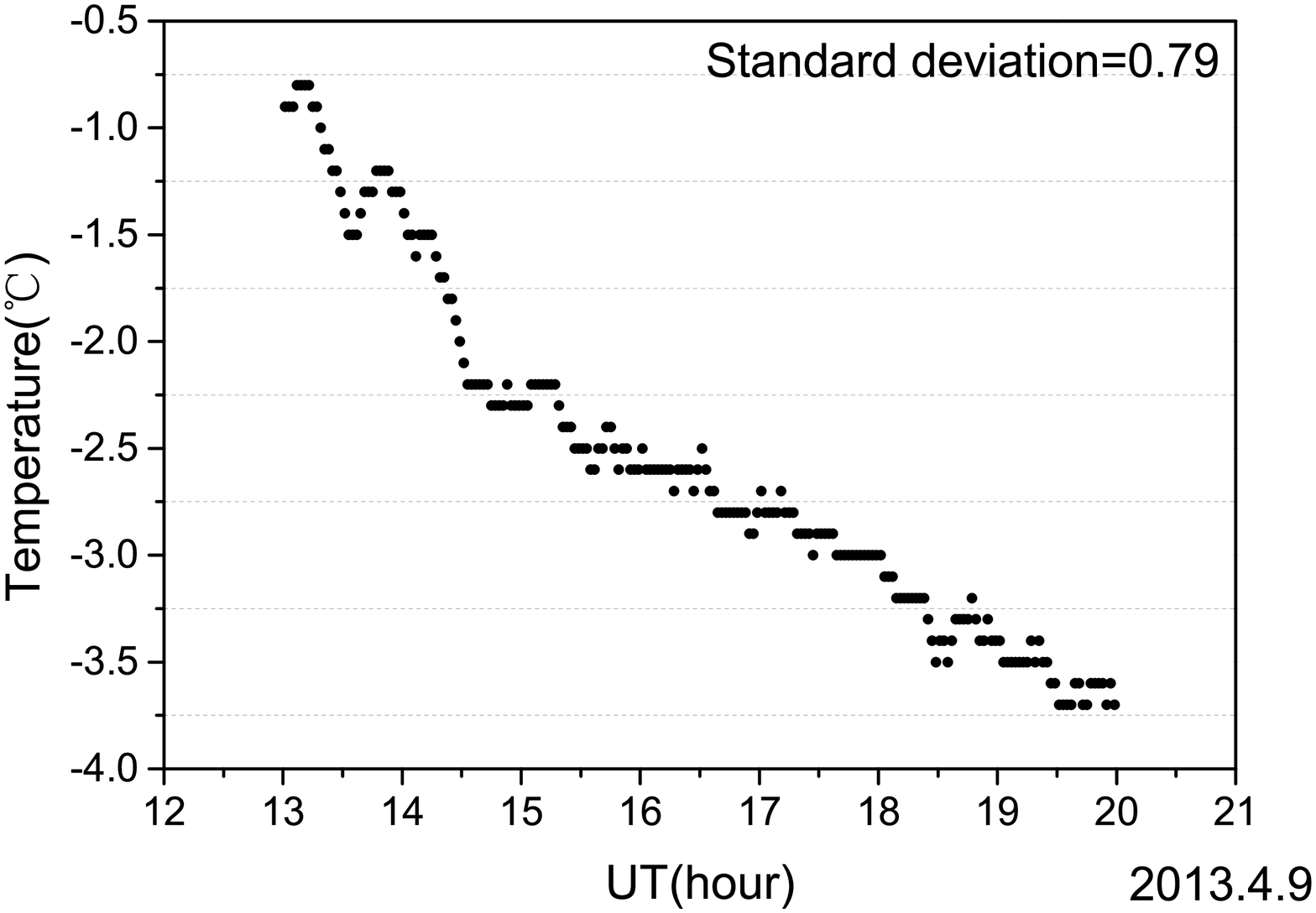}
\includegraphics[angle=0,scale=0.25]{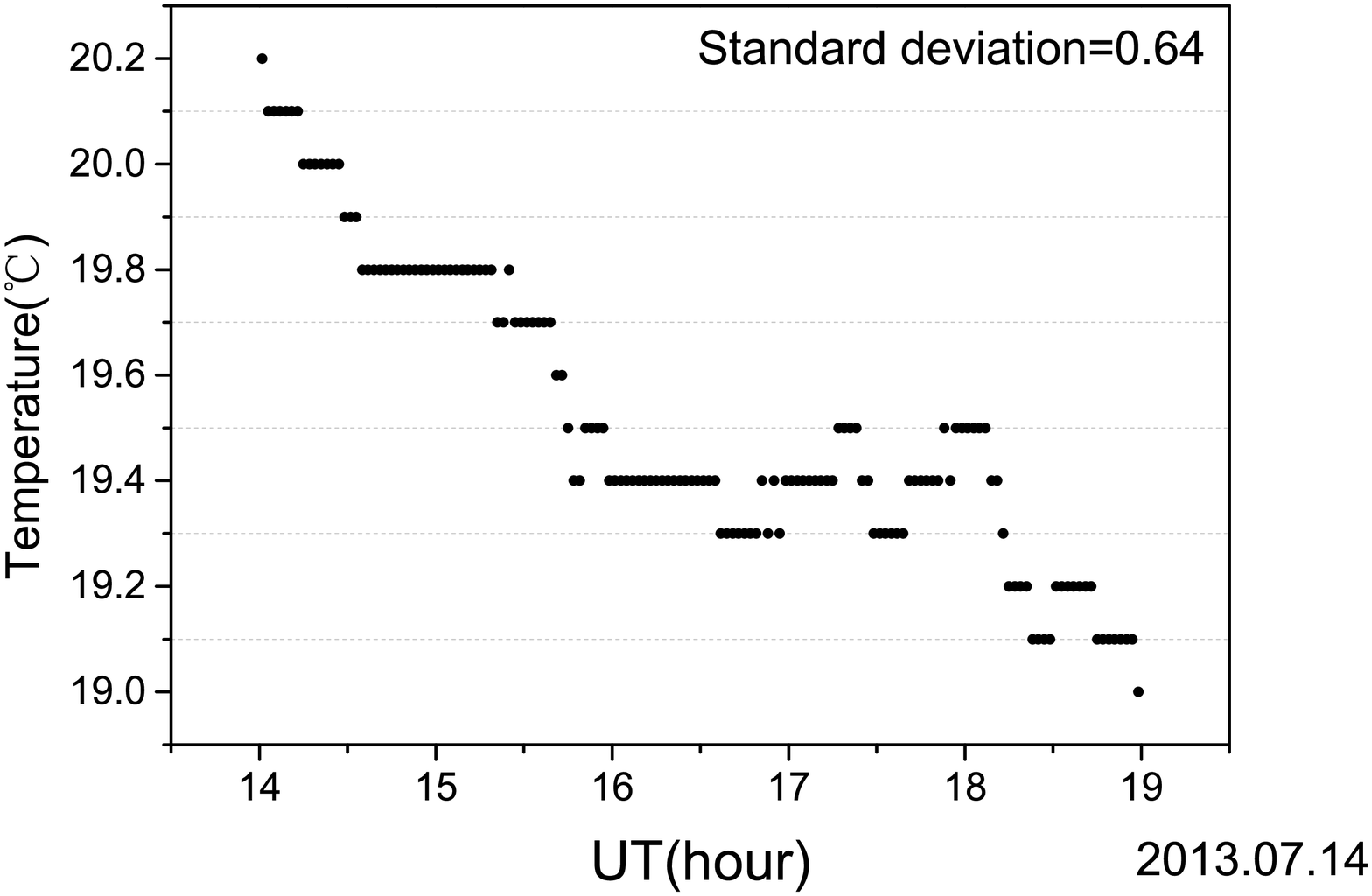}

\includegraphics[angle=0,scale=0.25]{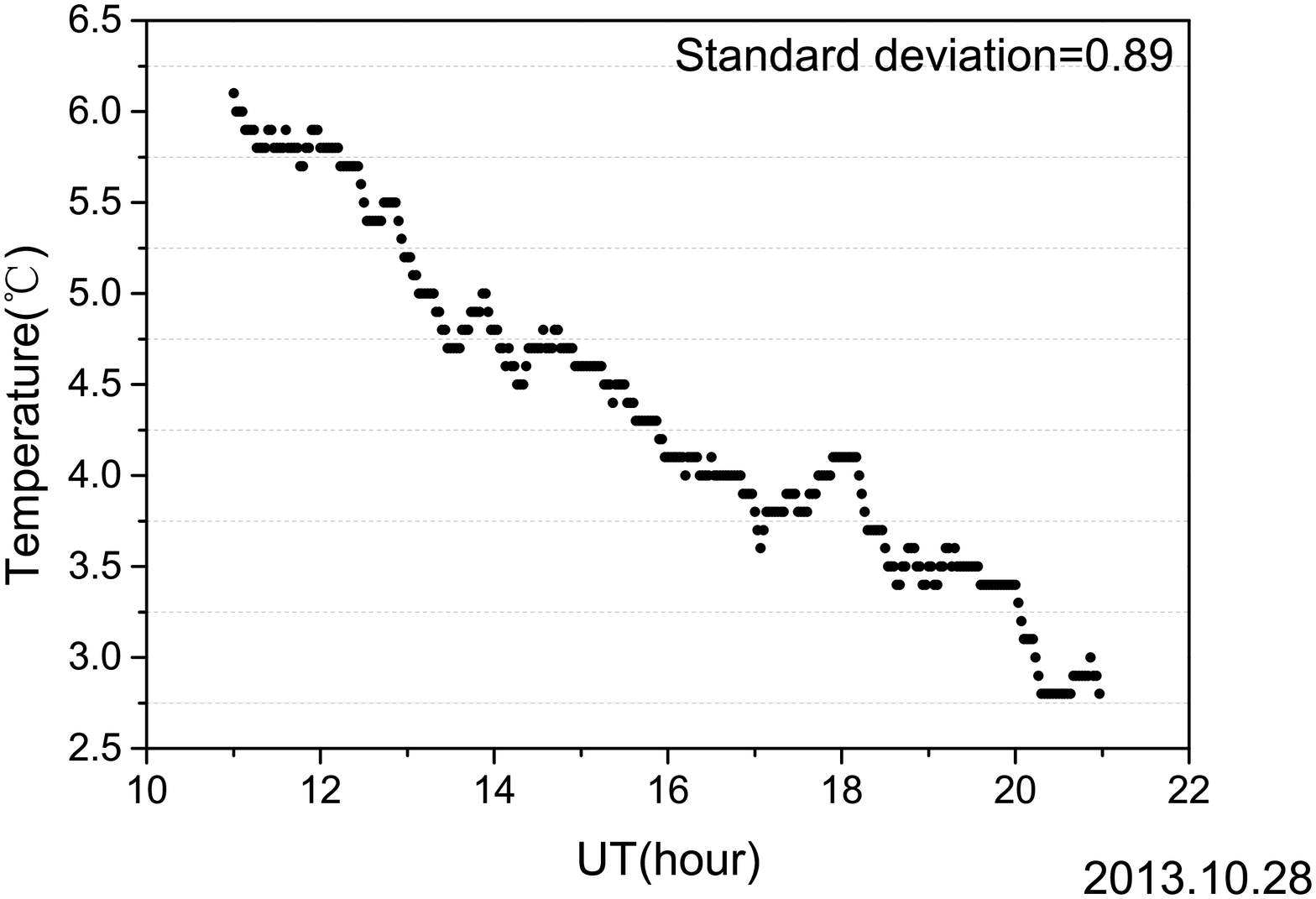}
\includegraphics[angle=0,scale=0.25]{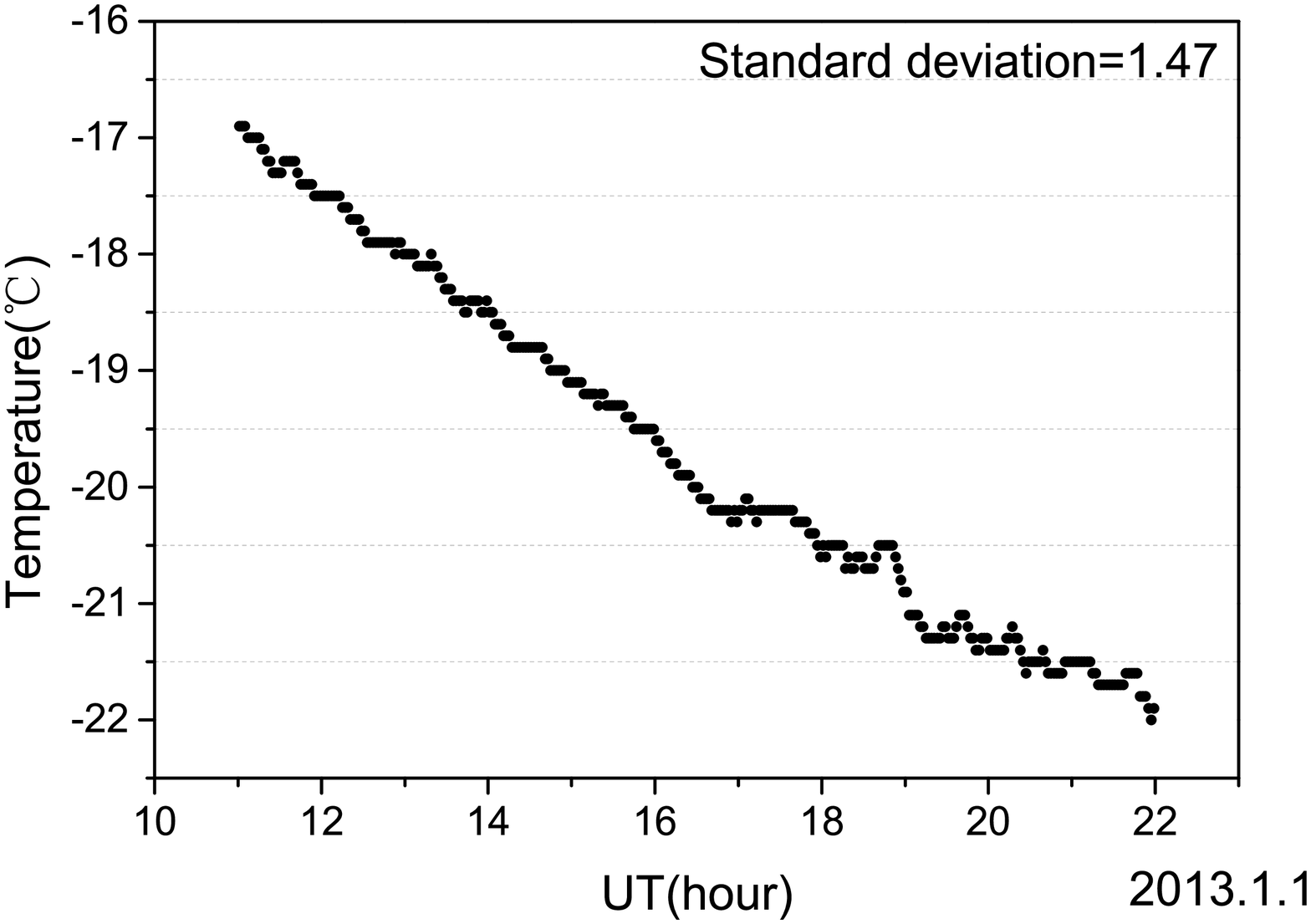}
\hfill \caption{Air temperature trend examples during nighttime of four seasons, spring, summer, autumn, and winter are shown with time, respectively. The time indicated in UT. Standard deviation are noted.}
\label{fig3}
\end{figure*}

\begin{figure*}[htbp]
\centering
\includegraphics[angle=0,scale=0.3]{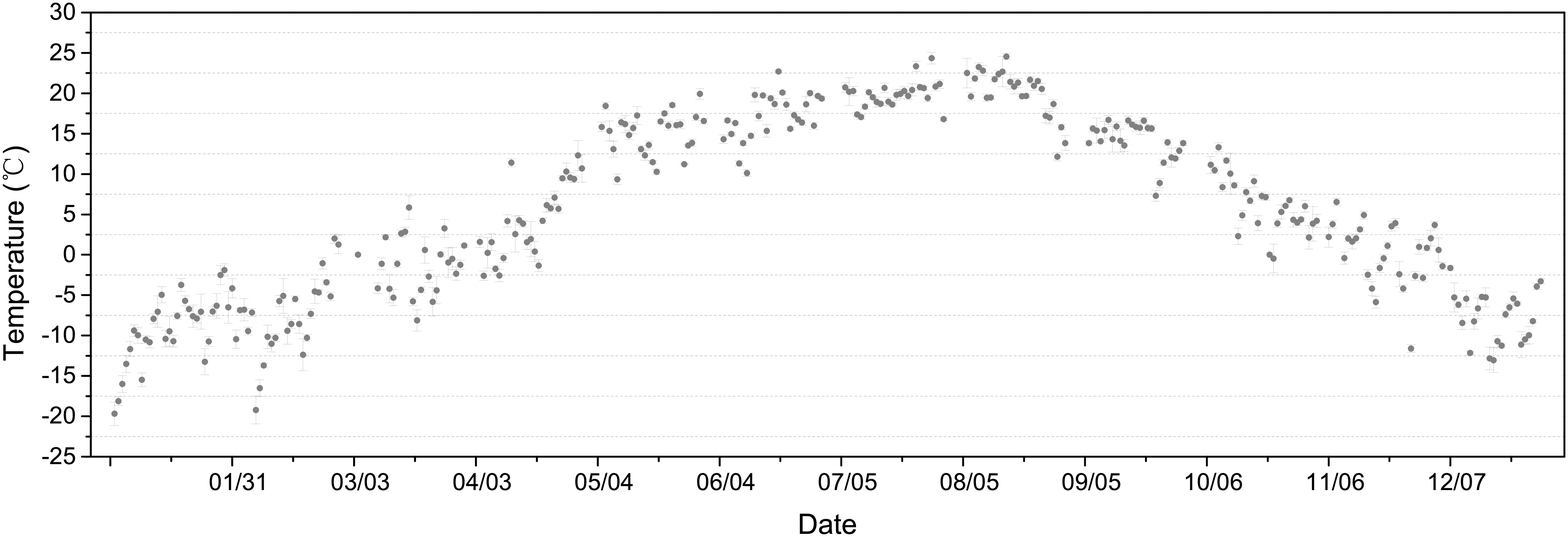}
\hfill \caption{Daily Mean of air temperature during nighttime and its standard deviation for the whole year of 2013 are shown.}
\label{fig4}
\end{figure*}

The airflow due to temperature difference influence the stability of atmosphere in the dome, which degrades the quality of images. In order to know the stability of the air temperature along the night, we select air temperature data between astronomical twilight of four nights from different seasons as an example. We have plotted the air temperature and its standard deviation with universal time (UT), Figure~\ref{fig3} shows that air temperature gradients are almost stable during the nighttime. Also, we have collected nighttime temperature data for whole year in 2013 as an example, and analyzed based on daily mean and its standard deviation, which are presented in Figure \ref{fig4}. The analysis suggested that the nighttime temperature at Xinglong observatory is almost stable, which indicates the site is suitable for astronomical observation.

\subsection{Wind Speed and Its Direction}

Wind speed is crucial in performing astronomical observations and telescope operation, as it influences the support structure and driver system of telescopes. As a safety precaution, observations are not allowed if the wind speed exceeds the defined upper safety limit, the slit of dome should be closed and the telescope should be back to the parking position. Various safety limits of wind speed at different sites are mentioned in \cite{2012MNRAS.422.2262R}, 15 m s$^{-1}$ is the typical maximum safety limit of the meteorological study in \cite{1985VA.....28..449M}. According to the local climate, Xinglong observatory follows 15 m s$^{-1}$ as the safety limits of wind speed.

\begin{figure*}[htbp]
\centering
\includegraphics[angle=0,scale=0.35]{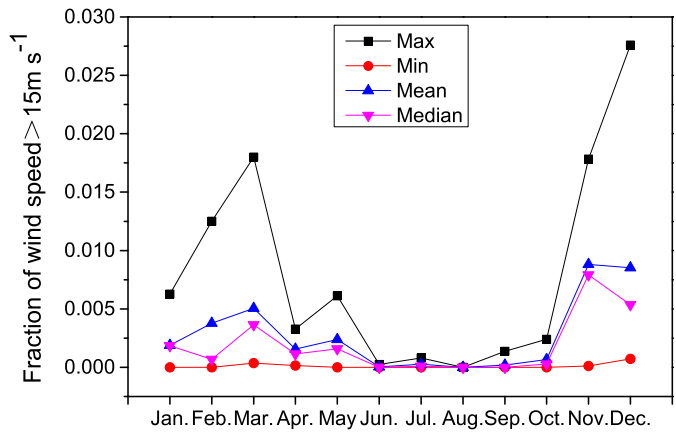}
\hfill \caption{Monthly statistics of wind speed above 15m s$^{-1}$ from 2007 to 2014 are shown. Note the major fraction occurs during winter. Black squares, red circles, blue triangles, and carmine inverted triangles indicate the maximum, minimum, mean, and median fractional value. See the electronic edition of the PASP for a color version of this figure.}
\label{fig5}
\end{figure*}

In order to understand the influence of wind speed and wind gusts on astronomical observation and telescope operation, we have obtained the data of wind speed using AWS from 2007 to 2014. Such long period of 8 years data allowed us to analyse the effect of wind speed on daytime, nighttime, and also 24 hours basis. The time criterion for nighttime data analysis is between astronomical twilight. We have ignored the $''$spurious data$''$ (the data points that are abnormal and far away from the adjacent data points) in our analysis. We have obtained the monthly distribution of nighttime wind speed above 15 m s$^{-1}$ and measured the maximum, minimum, mean and median value. We have performed the analysis for all the seasons, shown in Figure~\ref{fig5}, and found that the fraction of wind speed above 15 m s$^{-1}$ in winter (especially in November and December) and spring have more influence on observing time. These percentages implies the lost observing time due to high wind speed. We have noticed that the recorded maximum wind speed is around 27.6 m s$^{-1}$ (this value is instantaneous), which is important to be noted for further precautions and improvement of the facility.

\begin{figure*}[htbp]
\centering
\includegraphics[angle=0,scale=0.24]{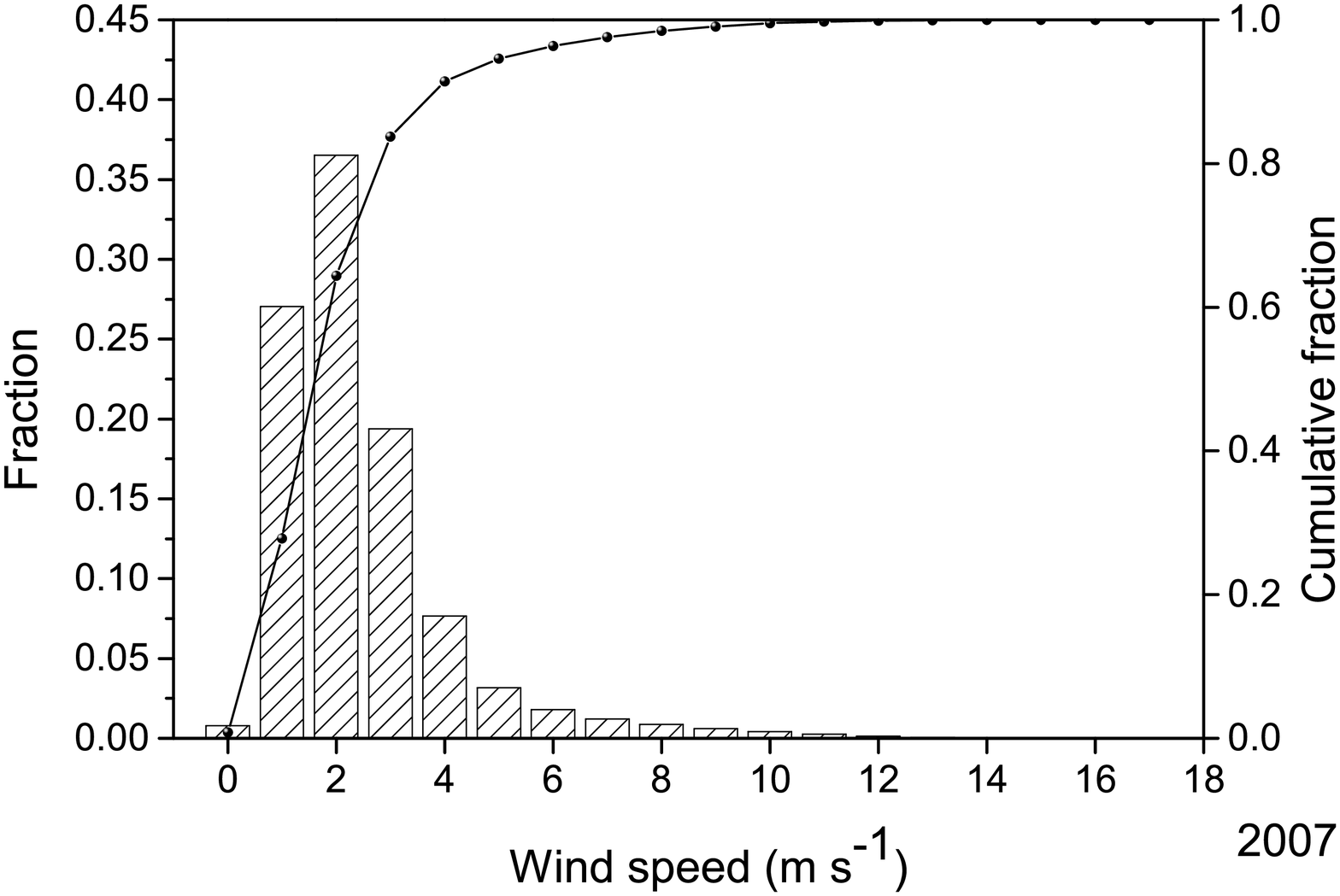}
\includegraphics[angle=0,scale=0.24]{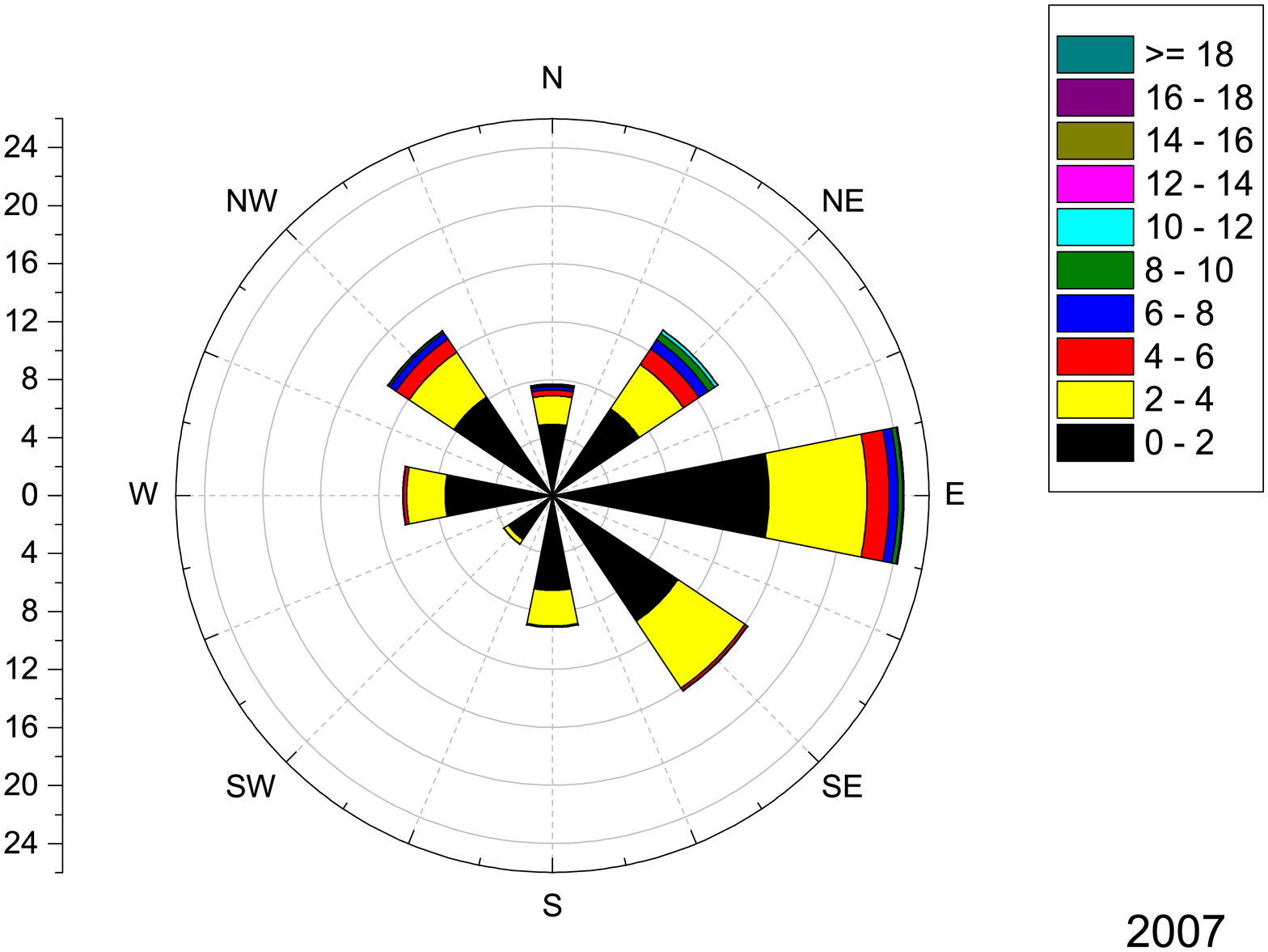}

\includegraphics[angle=0,scale=0.24]{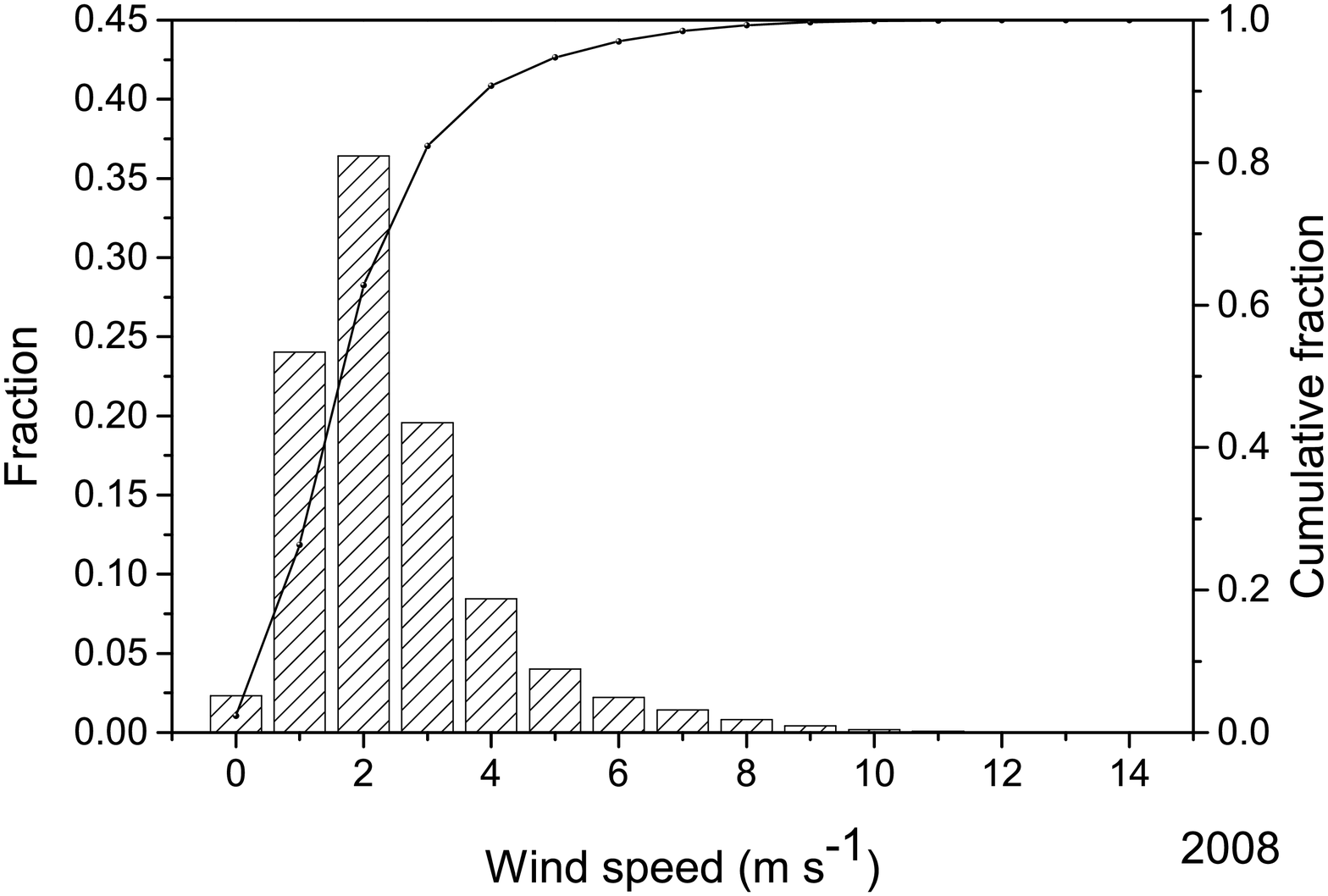}
\includegraphics[angle=0,scale=0.24]{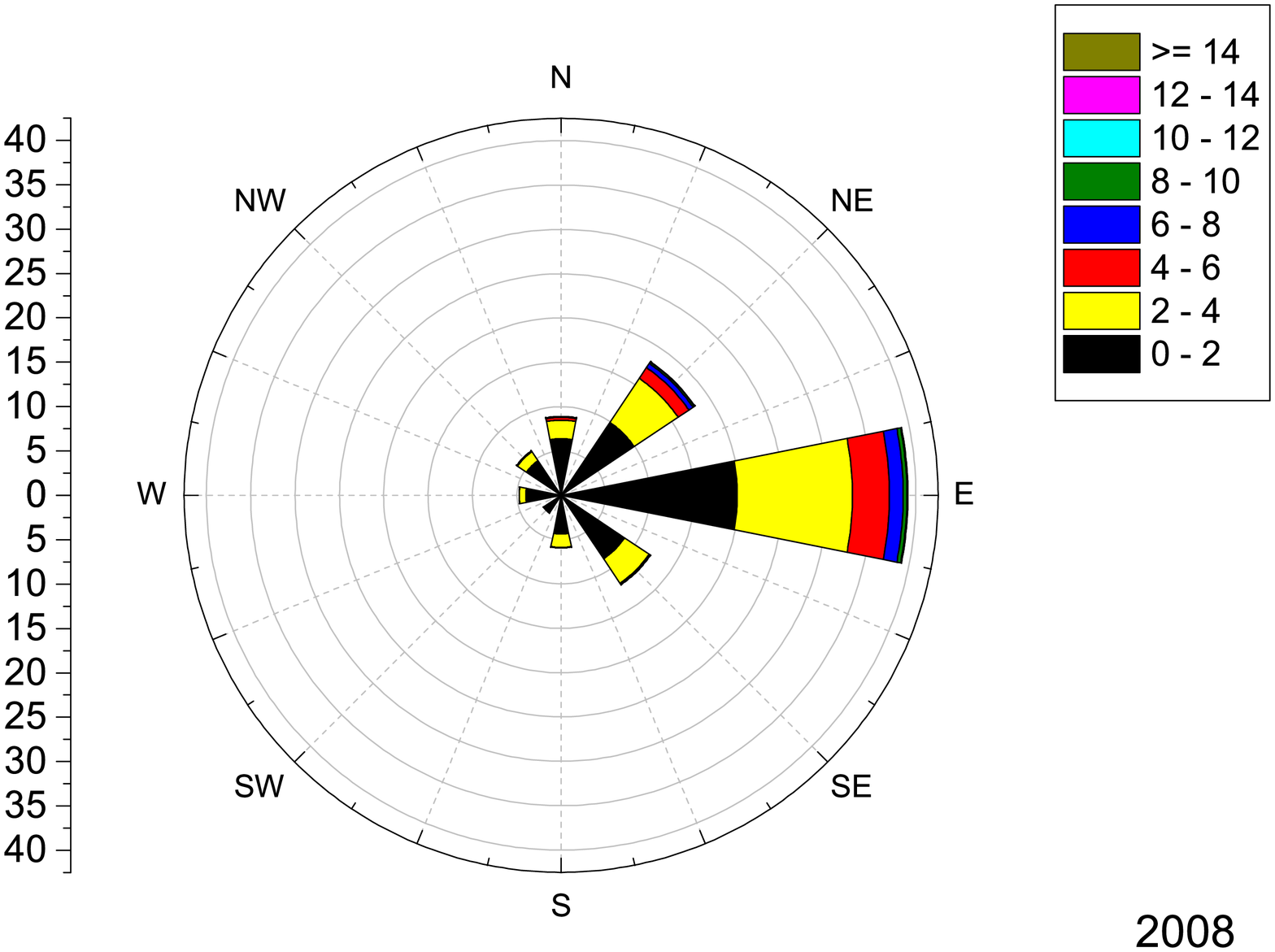}

\includegraphics[angle=0,scale=0.24]{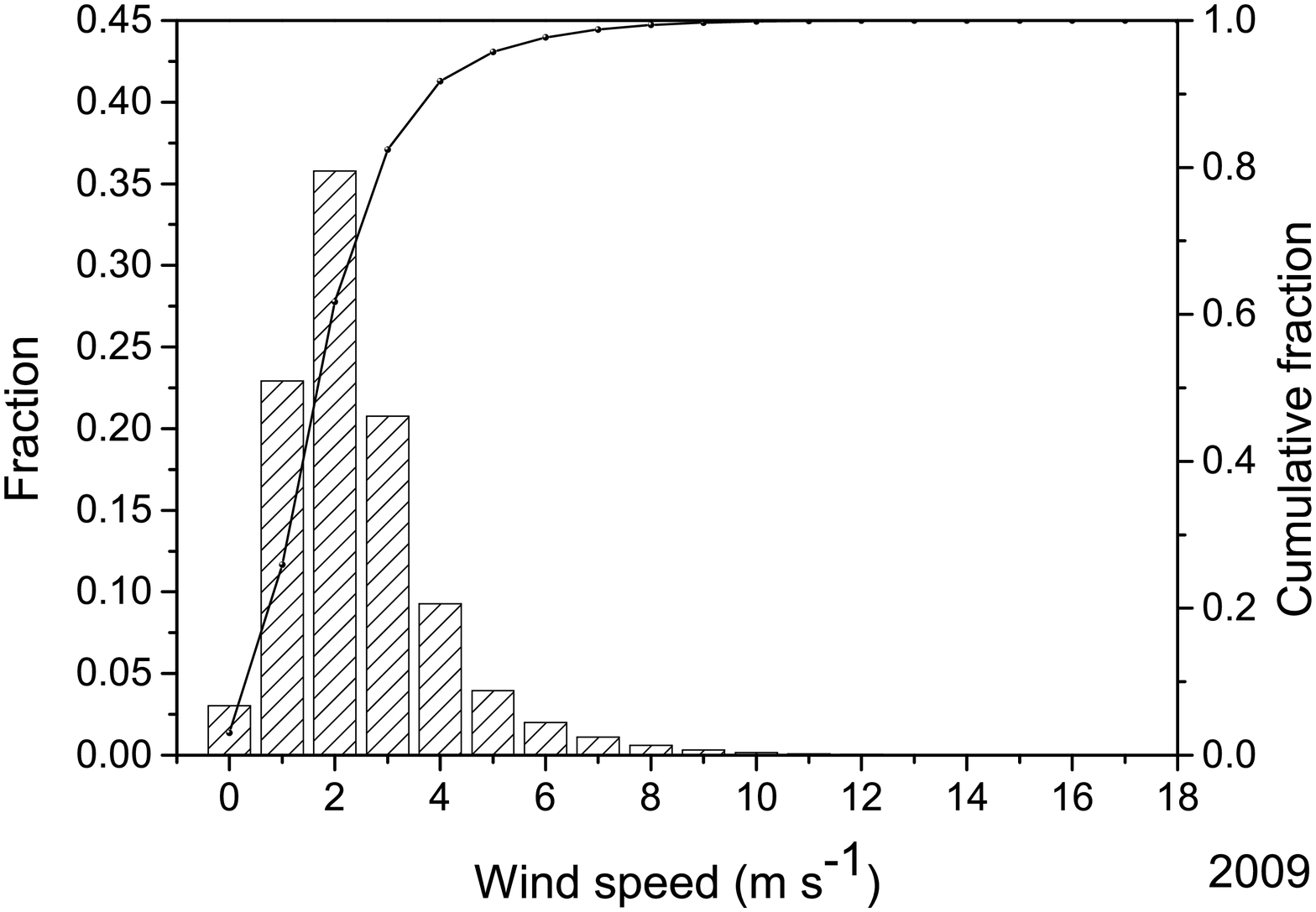}
\includegraphics[angle=0,scale=0.24]{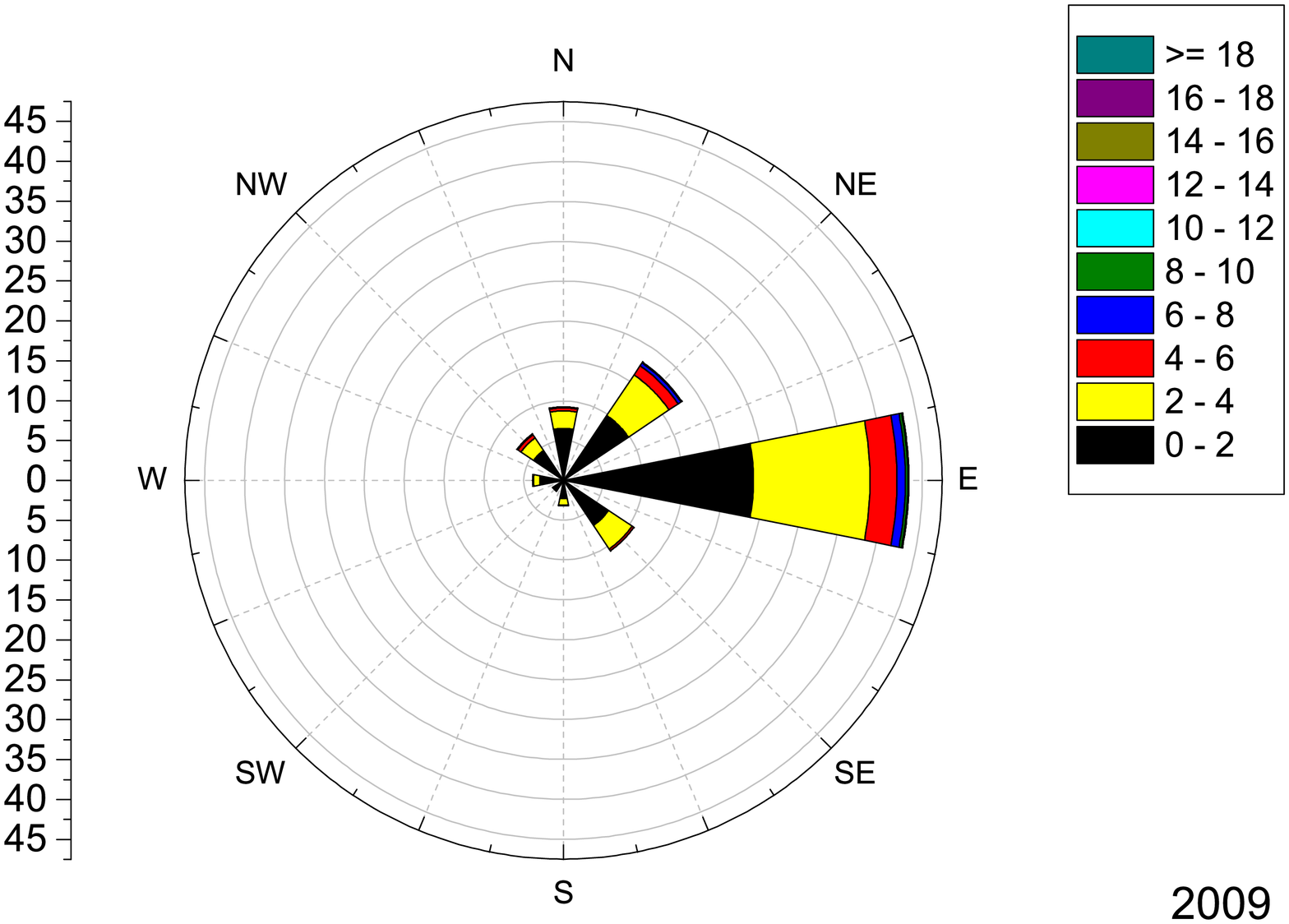}

\includegraphics[angle=0,scale=0.24]{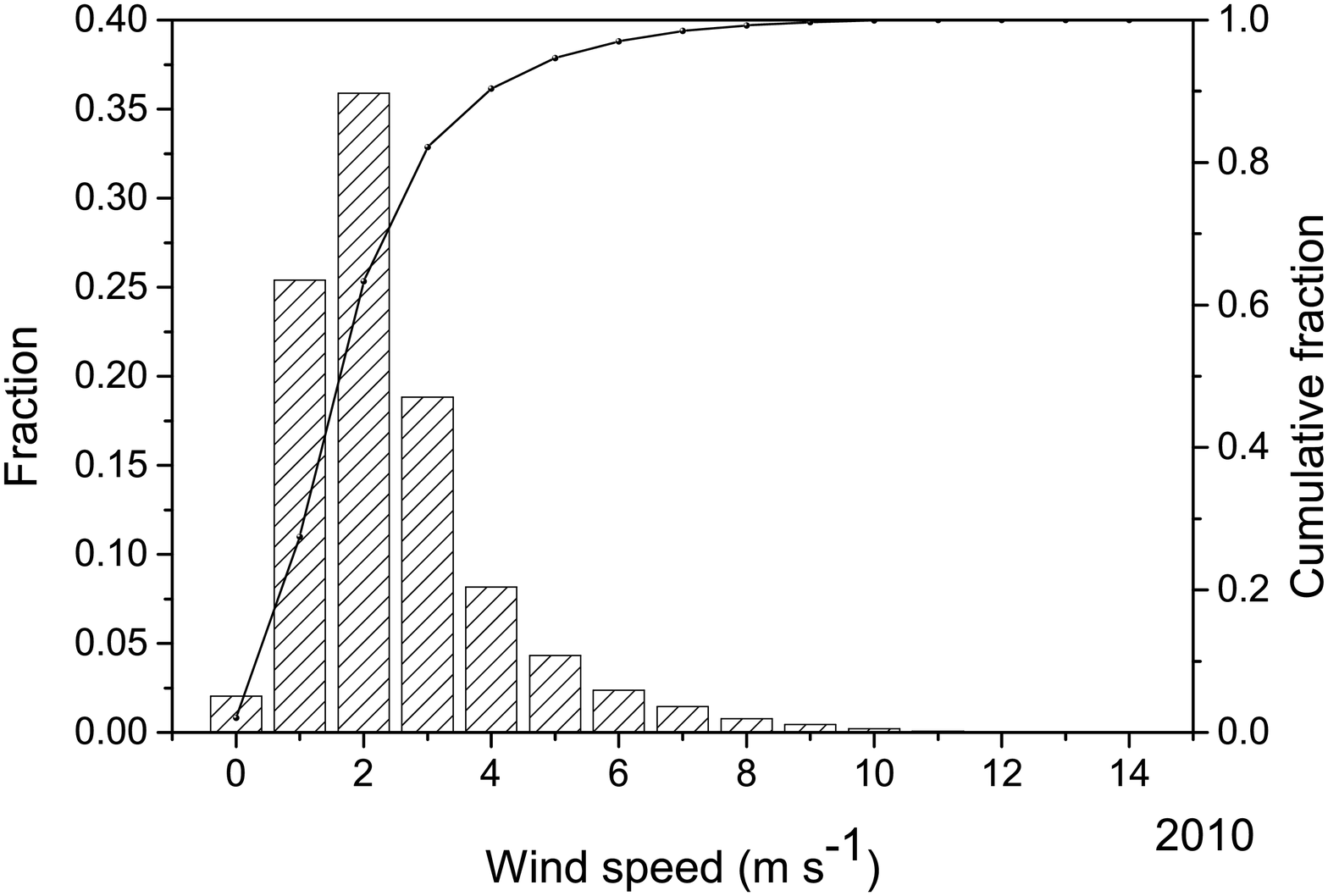}
\includegraphics[angle=0,scale=0.24]{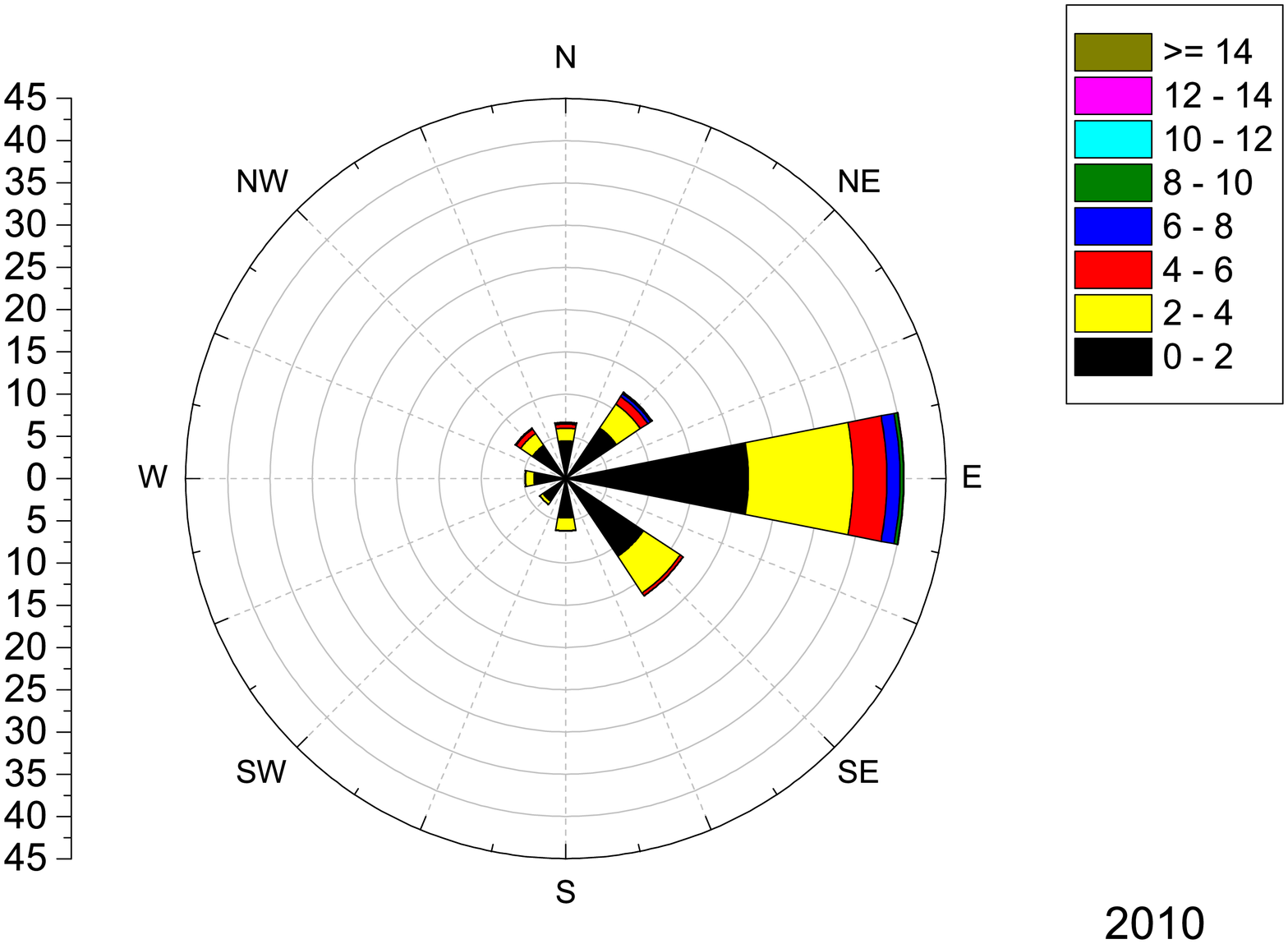}

\hfill \caption{Annual cumulative distribution of wind speed from 2007 to 2010 are shown in left panels. Fractional values and cumulative fraction of wind speed are indicated in left and right y-axis, respectively. Annual wind direction and frequency distributions are shown in right panels. The axis in left represents the frequency that corresponds to different directions, and the label in right represents the wind speed interval. See the electronic edition of the PASP for a color version of this figure.}
\label{fig6}
\end{figure*}

\begin{figure*}[htbp]
\centering
\includegraphics[angle=0,scale=0.24]{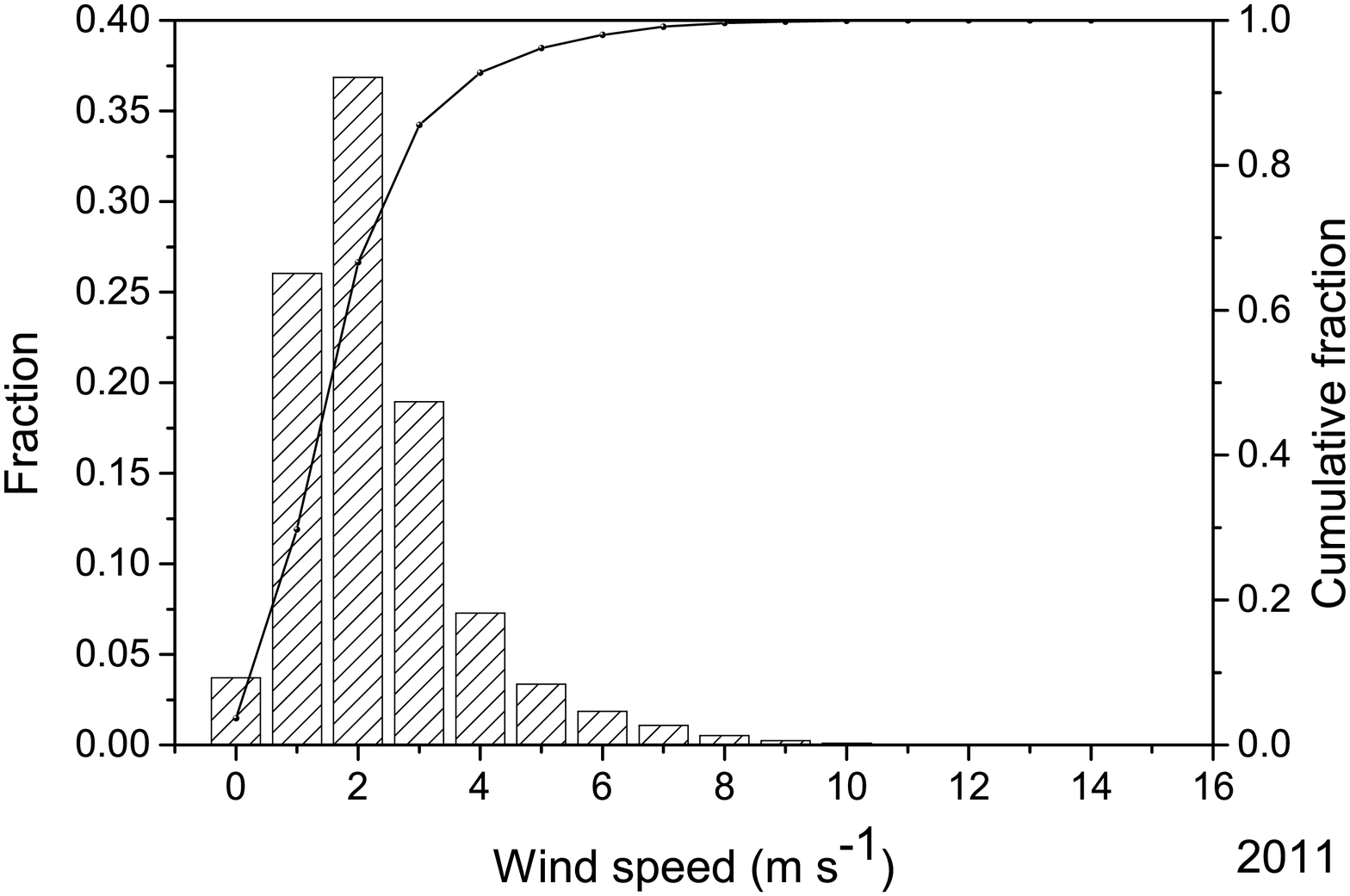}
\includegraphics[angle=0,scale=0.24]{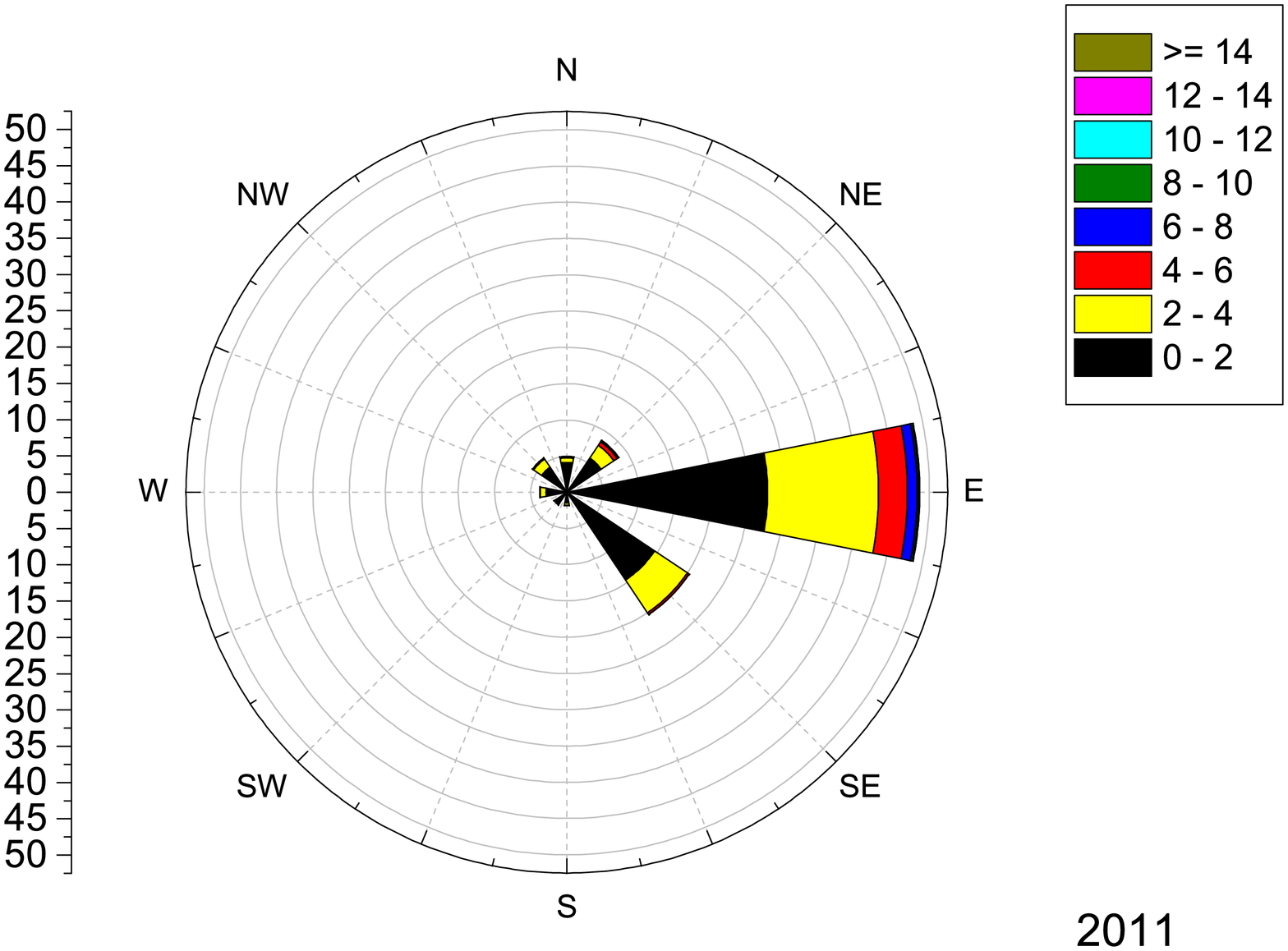}

\includegraphics[angle=0,scale=0.24]{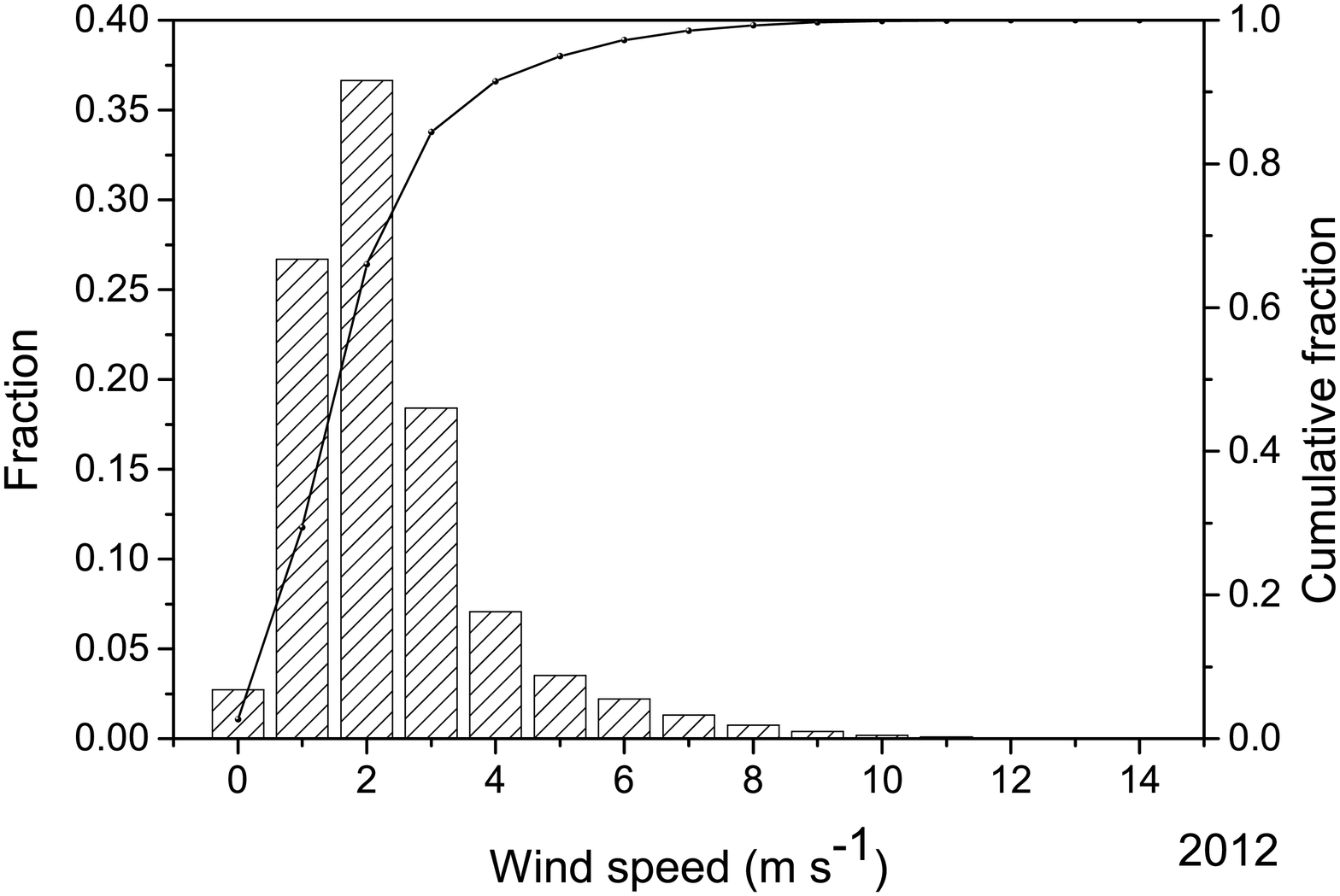}
\includegraphics[angle=0,scale=0.24]{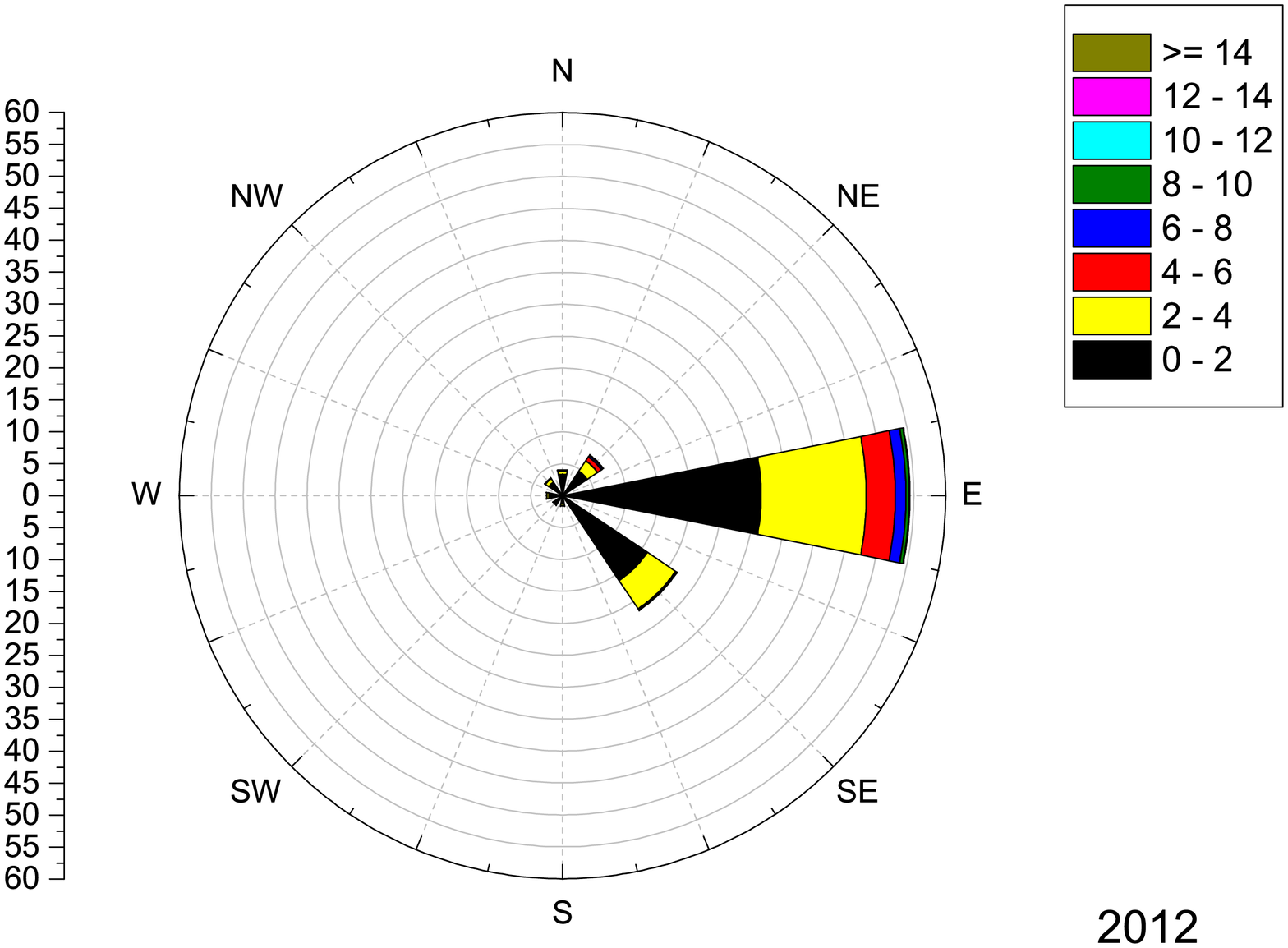}

\includegraphics[angle=0,scale=0.24]{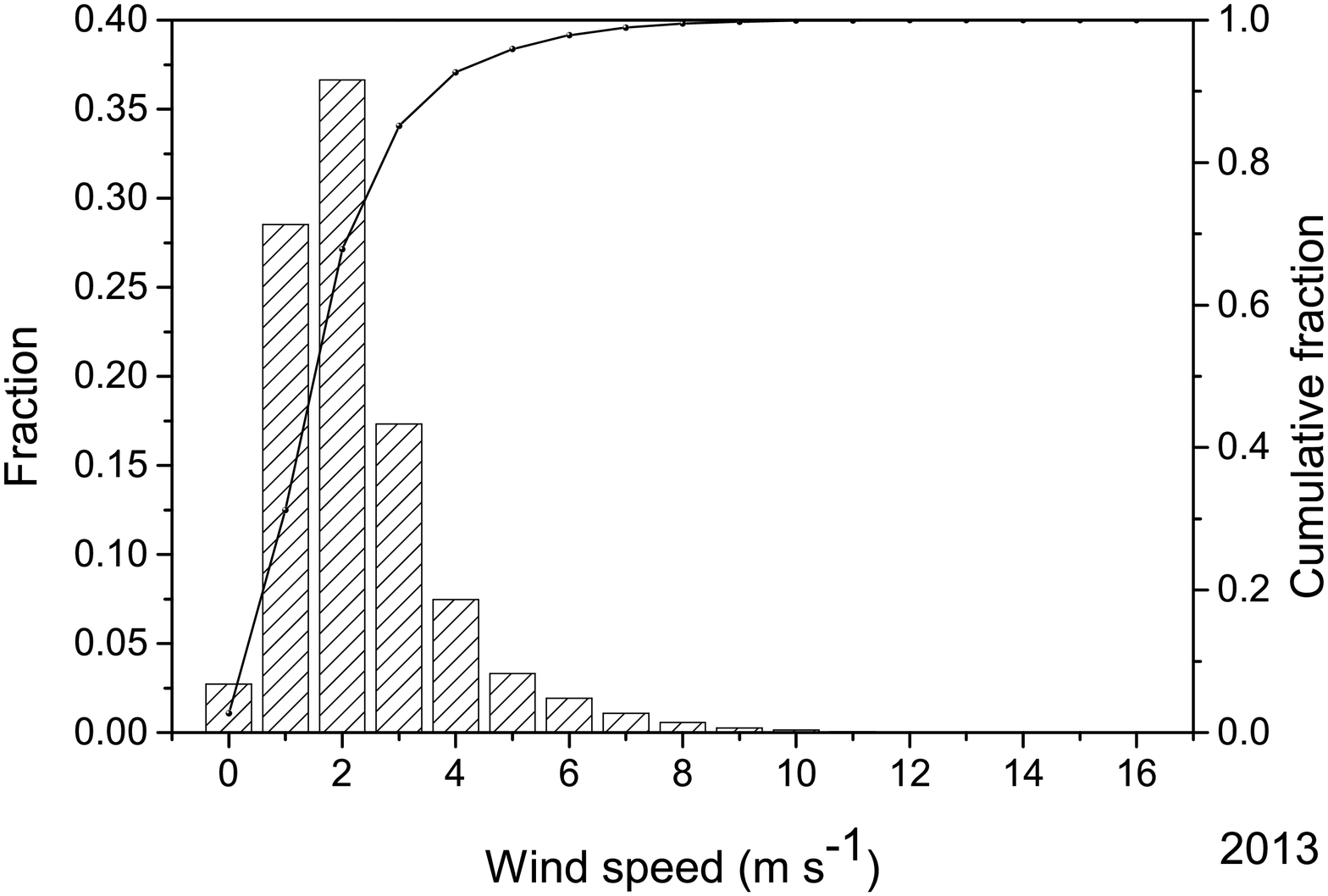}
\includegraphics[angle=0,scale=0.24]{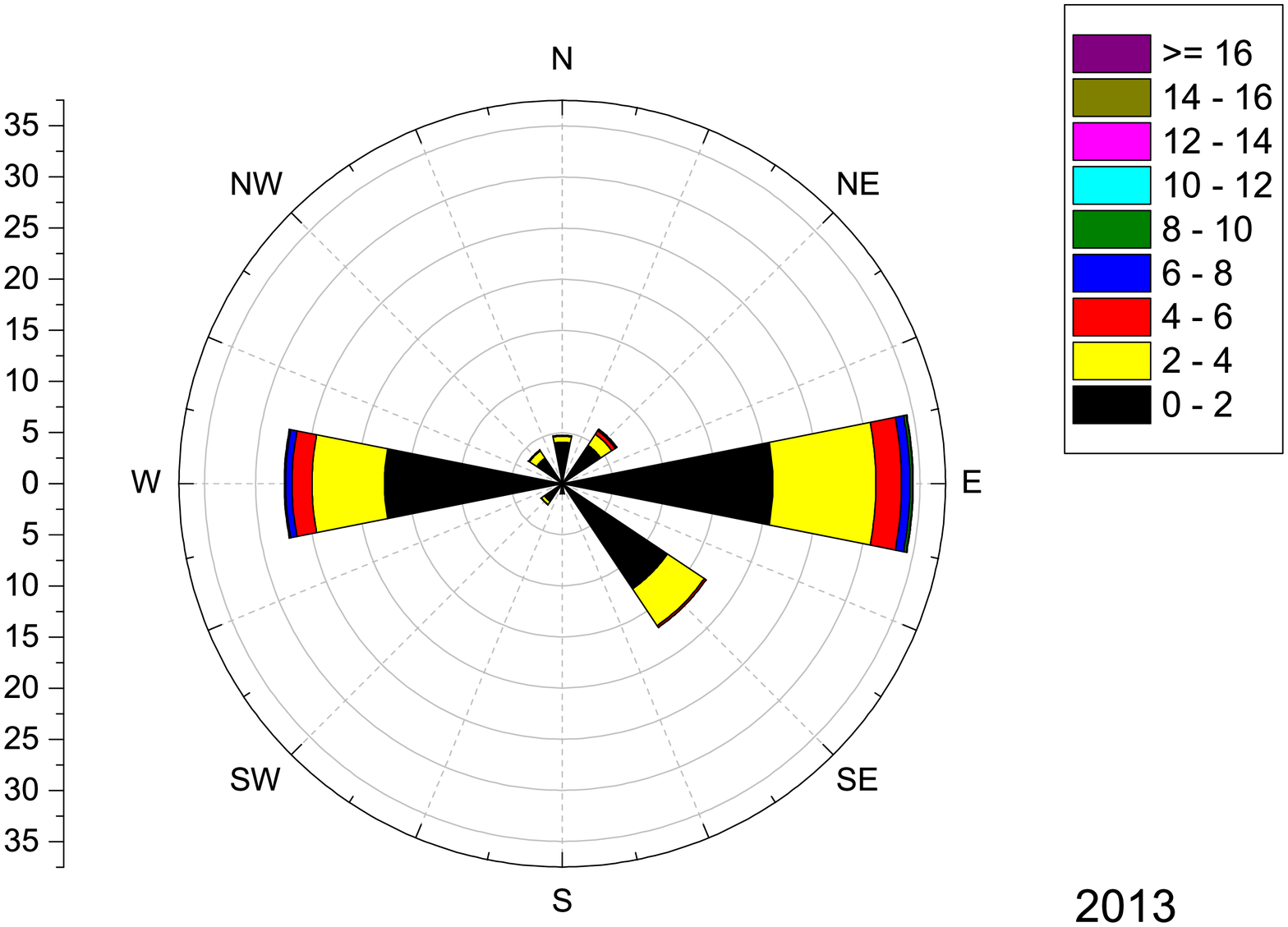}

\includegraphics[angle=0,scale=0.24]{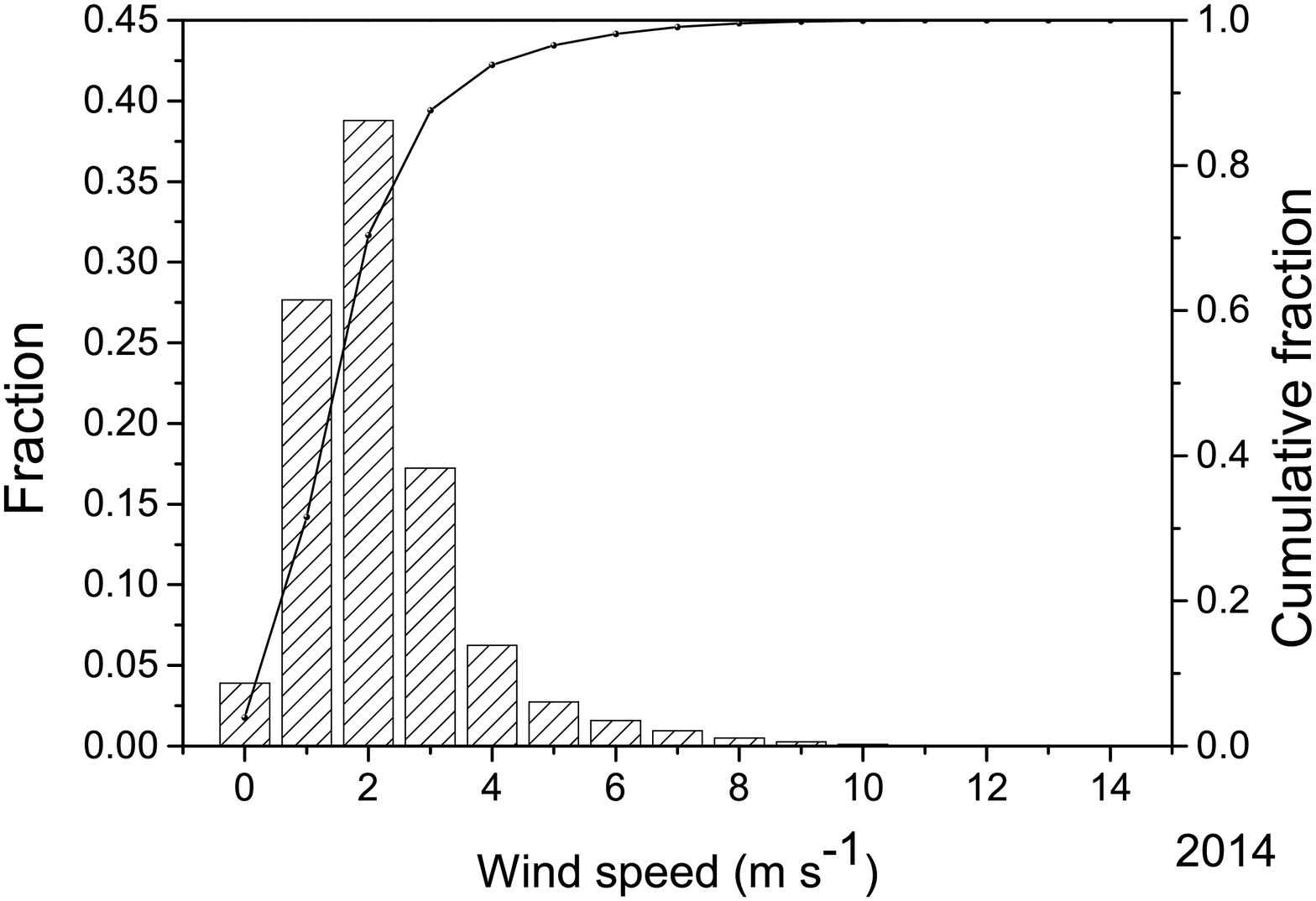}
\includegraphics[angle=0,scale=0.24]{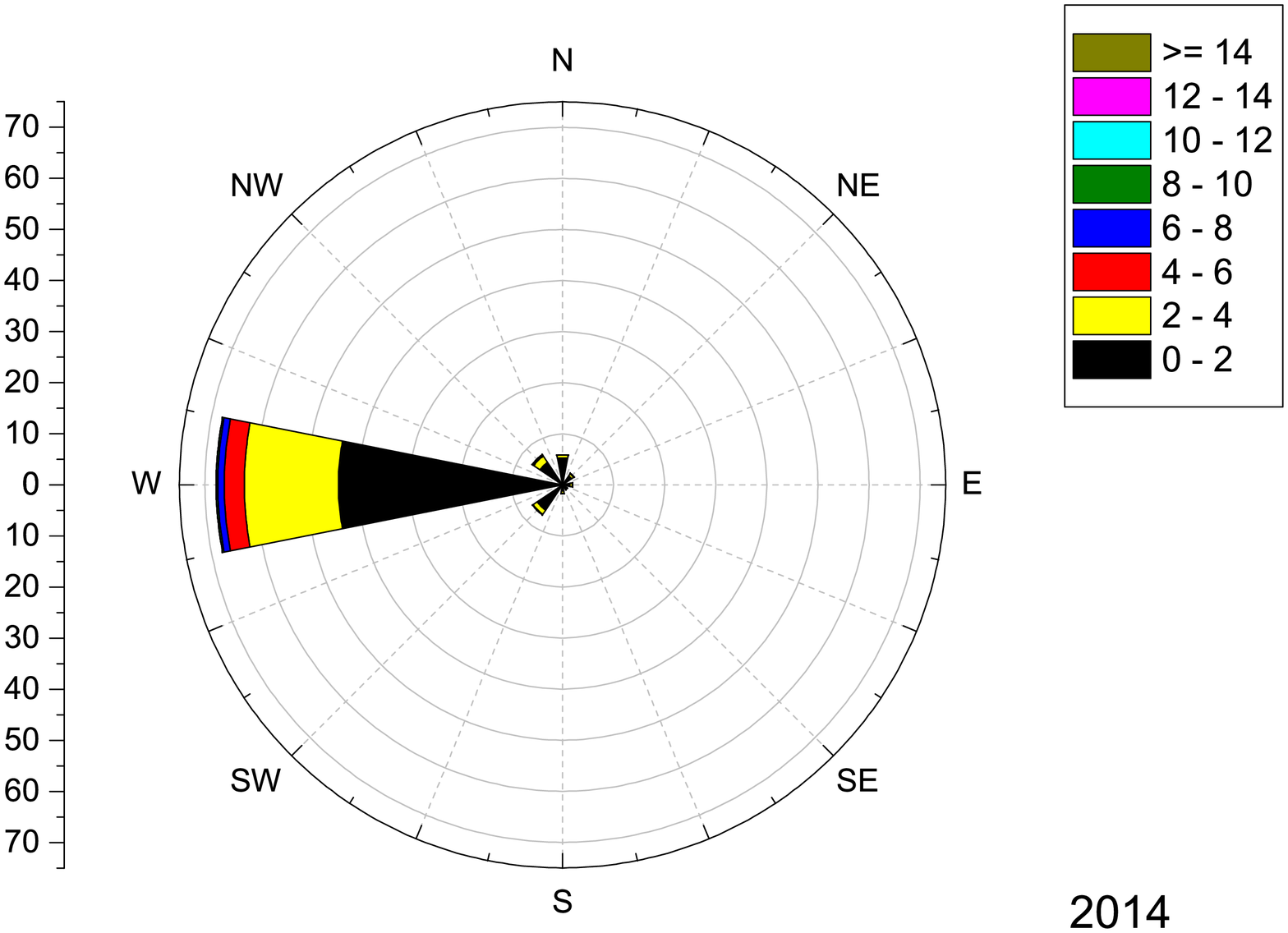}

\hfill \caption{Annual cumulative distribution of wind speed from 2011 to 2014 are shown in left panels. Fractional values  and cumulative fraction of wind speed are indicated in left and right y-axis, respectively. Annual wind direction and frequency distributions are shown in right panels.  The axis in left represents the frequency that corresponds to different directions, and the label in right represents the wind speed interval. See the electronic edition of the PASP for a color version of this figure.}
\label{fig7}
\end{figure*}

We have analysed the data of annual wind speed and its direction from 2007 to 2014 and shown in Figure \ref{fig6} and Figure \ref{fig7}. We find that most of the time wind speed is under 4~m s$^{-1}$, the peak value of distribution is around 2~m s$^{-1}$ during these years. Wind directions are mainly concentrated on East and West, which is influenced by the monsoon climate. We don't find other obvious regularity on the wind direction. Here we have summarized the monthly maximum, median, mean and its standard deviation of wind speed with 24hours, daytime and nighttime in Table 1,2 and 3, respectively. Notice that the mean and median wind speed are almost constant during the period analysed and ranged from 1.0 m s$^{-1}$ to 3.5 m s$^{-1}$, which are encouraging values for astronomical observations.

\begin{table}[htbp]
  \centering
  \caption{Monthly statistics on 24hrs data interval of wind speed and air temperature during 2007-2014}
%\footnotesize
  \setlength{\tabcolsep}{3pt}
    \begin{tabular}{ c c c c c c c c c c c c c c c c c c c c c  }
      \hline
      &&&Temperature& &\ \ \ \ &&Wind&& \\
      &&&($^\circ$C)& &\ \ \ &&(m s$^{-1}$)&& \\
      \hline
Month    & Ndata & Max    & Min   & Median &  Mean    & Std      && Max    &Median &Mean  &Std\\
      \hline
January  & 169641   &  8.9   &-22.0  & -8.2   &  -8.18   &   4.70   &&  23.4   &1.8   &2.25  &2.01\\
February & 158937   & 11.1   &-21.7  & -4.9   &  -4.72   &   5.35   &&  24.5   &1.8   &2.40  &2.22\\
March    & 174292   & 22.6   &-15.3  &  0.9   &  1.33    &   5.72   &&  24.4   &2.3   &3.03  &2.63\\
April    & 168748   & 26.1   &-5.3   &  8.8   &  8.88    &   5.32   &&  23.4   &2.4   &3.08  &2.45\\
May      & 174205   & 33.4   &0.6    & 16.1   &  16.23   &   4.90   &&  27.6   &2.3   &2.94  &2.43\\
June     & 160408   & 31.4   &9.2    & 19.2   &  19.38   &   4.03   &&  21.6   &1.6   &2.02  &1.58\\
July     & 155463   & 33.7   &12.8   & 21.7   &  22.02   &   3.27   &&  23.0   &1.4   &1.78  &1.31\\
August   & 151235   & 32.7   &11.5   & 20.6   &  20.94   &   3.37   &&  16.0   &1.4   &1.69  &1.28\\
September& 166691   & 29.5   &3.8    & 16.0   &  16.16   &   4.07   &&  18.1   &1.4   &1.79  &1.45\\
October  & 160894   & 23.4   &-3.9   & 10.0   &  9.94    &   4.58   &&  20.5   &1.6   &2.18  &1.92\\
November & 167998   & 20.2   &-14.1  &  1.2   &  0.95    &   5.08   &&  25.7   &1.8   &2.41  &2.28\\
December & 170232   & 12.3   &-21.6  & -6.4   &  -6.68   &   5.06   &&  25.0   &2.0   &2.66  &2.49\\
     \hline
    \end{tabular}
\end{table}

\begin{table}[htbp]
  \centering
  \caption{Monthly statistics on daytime data of wind speed and air temperature during 2007-2014}
%\footnotesize
  \setlength{\tabcolsep}{3pt}
     \begin{tabular}{ c c c c c c c c c c c c c c c c c c c c c  }
      \hline
      &&&Temperature& &\ \ \ \ &&Wind&& \\
      &&&($^\circ$C)& &\ \ \ &&(m s$^{-1}$)&& \\
	  \hline
Month     & Ndata  & Max    & Min    & Median  & Mean    & Std     && Max    &Median &Mean & Std\\
      \hline
January   & 91288  &  8.9  & -21.9   &   -7.1  &  -7.20  &  4.81   && 23.4   &1.8   &2.37  &2.07\\
February  & 90576  & 11.1  & -21.7   &   -3.7  &  -3.64  &  5.33   && 24.5   &2.0   &2.60  &2.25\\
March     & 108148 & 22.6  & -15.3   &    2.0  &   2.28  &  5.88   && 24.4   &2.5   &3.24  &2.70\\
April     & 116691 & 26.1  &  -5.3   &    9.8  &   9.77  &  5.39   && 23.4   &2.6   &3.31  &2.53\\
May       & 132085 & 33.4  &   0.7   &   17.0  &  17.01  &  4.94   && 27.6   &2.4   &3.11  &2.54\\
June      & 126805 & 31.4  &   9.2   &   19.9  &  19.93  &  4.12   && 21.6   &1.7   &2.12  &1.64\\
July      & 122410 & 33.7  &  13.0   &   22.4  &  22.52  &  3.30   && 23.0   &1.5   &1.86  &1.35\\
August    & 107450 & 32.7  &  11.5   &   21.5  &  21.71  &  3.47   && 16.0   &1.5   &1.80  &1.30\\
September & 107362 & 29.5  &   4.5   &   17.1  &  17.14  &  4.16   && 18.1   &1.5   &1.89  &1.46\\
October   & 94919  & 23.4  &  -3.1   &   11.1  &  10.94  &  4.73   && 20.5   &1.8   &2.34  &2.03\\
November  & 90875  & 20.2  & -13.5   &    2.2  &   1.97  &  5.14   && 25.7   &1.9   &2.55  &2.39\\
December  & 92944  & 12.3  & -21.2   &   -5.6  &  -5.83  &  5.08   && 25.0   &2.0   &2.73  &2.51\\
     \hline
    \end{tabular}
\end{table}

\begin{table}[htbp]
  \centering
  \caption{Monthly statistics on nighttime data of wind speed and air temperature during 2007-2014}
%\footnotesize
  \setlength{\tabcolsep}{3pt}
    \begin{tabular}{ c c c c c c c c c c c c c c c c c c c c c c }
      \hline
      &&&Temperature& &\ \ \ &&Wind&& \\
      &&&($^\circ$C)& &\ \ &&(m s$^{-1}$)&& \\
	  \hline
Month    & Ndata  & Max  & Min    & Median &  Mean  & Std    && Max    & Median & Mean & Std\\
      \hline
January  & 78353   & 2.5  &  -22.0 & -9.4   & -9.30  & 4.31   && 19.6   & 1.7   &2.12  &1.91\\
February & 68361   & 6.7  &  -21.7 & -6.4   & -6.15  & 5.04   && 23.0   & 1.6   &2.15  &2.15\\
March    & 66144   & 17.3 &  -15.3 & -0.5   & -0.24  & 5.07   && 24.0   & 2.0   &2.70  &2.48\\
April    & 52057   & 19.5 &   -5.3 &  6.9   & 6.91   & 4.60   && 20.4   & 2.0   &2.57  &2.16\\
May      & 42120   & 27.4 &    0.6 & 14.1   & 13.82  & 3.90   && 20.9   & 1.9   &2.40  &1.96\\
June     & 33603   & 26.7 &    9.4 & 17.4   & 17.29  & 2.82   && 15.9   & 1.4   &1.66  &1.26\\
July     & 33053   & 26.5 &   12.8 & 20.1   & 20.18  & 2.44   && 19.3   & 1.2   &1.47  &1.10\\
August   & 43785   & 25.3 &   12.2 & 19.2   & 19.11  & 2.27   && 12.9   & 1.2   &1.42  &1.18\\
September& 59329   & 23.1 &    3.8 & 14.7   & 14.43  & 3.19   && 15.7   & 1.2   &1.59  &1.41\\
October  & 65975   & 17.1 &   -3.9 &  8.7   & 8.49   & 3.92   && 17.6   & 1.4   &1.94  &1.73\\
November & 77123   & 13.9 &  -14.1 &  0.0   & -0.25  & 4.73   && 23.3   & 1.7   &2.26  &2.14\\
December & 77288   & 9.5  &  -21.6 & -7.3   & -7.71  & 4.82   && 22.0   & 1.9   &2.58  &2.48\\
     \hline
    \end{tabular}
\end{table}

\subsection{Relative Humidity}

Moisture and water condensation are the extreme bad results from higher relative humidity which effects the astronomical observation and facilities. Condensation is a serious issue for not only telescope operation by damaging the light sensors and electronic equipment but also degrade the quality of observed astronomical data \citep{2012MNRAS.422.2262R}. Higher humidity allows dust particles on exposed air to settle down on mirrors, which will decrease their reflectivity. \cite{2000A&AS..147..271J} and \cite{2007PASP..119..292L} suggested that the observations should be stopped when the relative humidity goes beyond 90\%. \cite{2009MNRAS.399..783L} reported the safety limits on relative humidity between 80\% and 85\% for the Paranal Observatory, located on the coast of the Atacama Desert (Chile). After taking into consideration of the above studies, Xinglong observatory set safety limits according to the local climate, in which telescope operation is not allowed if the relative humidity exceeds 90\% before observation or exceeds 95\% during observation.

\begin{figure*}[htbp]
\centering
\includegraphics[angle=0,scale=0.3]{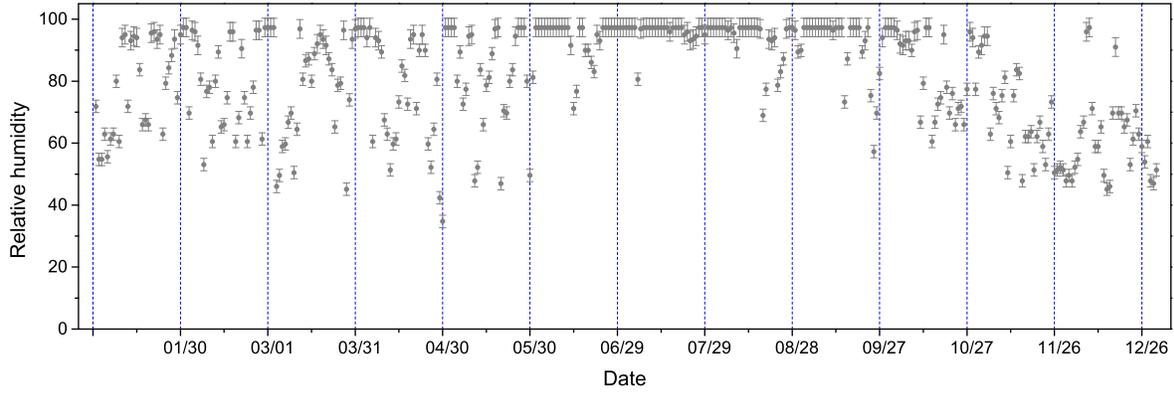}
\hfill \caption{Behaviour of relative humidity based on daily maximum for the whole year 2013 is shown. Error bars indicate the precision of measurements.}
\label{fig8}
\end{figure*}

\begin{figure*}[htbp]
\centering
\includegraphics[angle=0,scale=0.3]{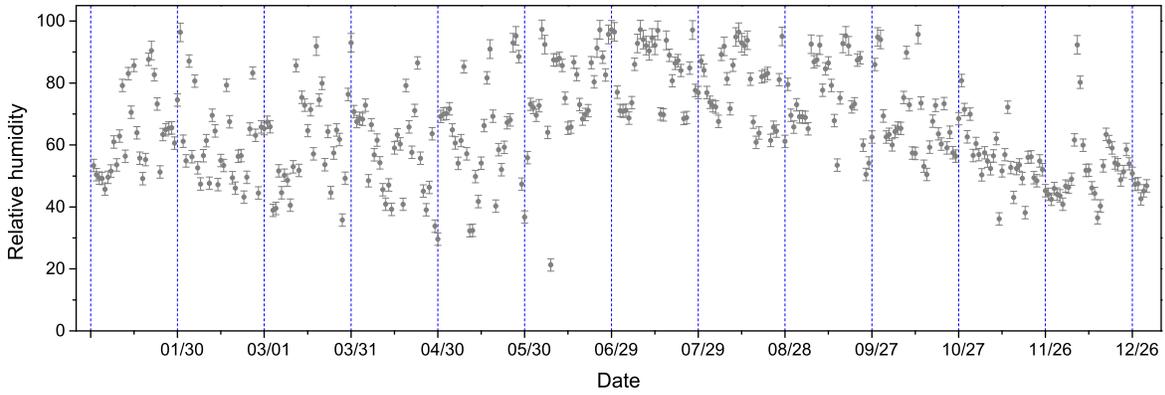}
\hfill \caption{Behaviour of relative humidity based on daily mean for the whole year 2013 is shown. Error bars indicate the precision of measurements.}
\label{fig9}
\end{figure*}

In order to understand the behaviour of humidity at Xinglong observatory, we have collected the relative humidity data from AWS for the whole year of 2013 and analysed the maximum and mean value on daily basis. Daily maximum and mean of relative humidity data are shown in Figure~\ref{fig8} and Figure~\ref{fig9}, respectively. The error bars indicates the precision of measurements. We have noticed that the large fraction of humid time occurs in summer and it is around 55\%, may be due to monsoon climate and rains. High humidity is one of the main factors for lost time in summer at Xinglong observatory.

\subsection{Photometric Nights and Spectroscopic Nights}

Definition of $''$photometric nights$''$ at Xinglong observatory is cloud free observations which last for at least 6 hours or whole night in summer as nights are shorter. $''$Spectroscopic nights$''$ means that observations are performed in both clear sky and partial cloudy conditions.

Data of photometric nights and spectroscopic nights are collected from the telescope observation logs wherein night assistants record the observing information of each night between 2007 and 2014. Note that the information in logs are judged mainly on visual experience and meteorological parameters during that time, and expected to have some uncertainties. We have gathered information from the observation logs of three telescopes (the 2.16-meter reflector, the 0.85-meter reflector, the 0.8-meter reflector) at Xinglong observatory and compared them through comprehensive analysis.

The annual statistics of photometric nights, spectroscopic nights, useful nights, and unuseful nights (nights that are not able to observe due to bad weather), from 2007 to 2014 are given in Table~4. Average fraction of photometric nights and spectroscopic nights are 32\% and 63\% per year, respectively. Number of useful nights are equal to the number of spectroscopic nights at Xinglong observatory. In addition, we have calculated the annual useful and unuseful nights. Average number of useful nights and unuseful nights are 230 and 135 per year, respectively. Fraction of photometric nights and spectroscopic nights are approximately equal during these years, which means the weather conditions of Xinglong observatory are almost constant and suitable for optical observations.

\begin{table}[htbp]
  \centering
%\footnotesize
\caption{Annual statistics of photometric nights, spectroscopic nights, useful time, and unuseful time during 2007-2014.}
    \begin{tabular}{ c c c c c }
    \hline
     Year&Photometric nights&Spectroscopic nights&Useful time&Unuseful time\\
     &(\%) & (\%) &&\\
      \hline
     2007 &	28 & 69 & 253 & 112 \\
     2008 &	31 & 60 & 220 & 145 \\
     2009 &	33 & 60 & 219 & 146 \\
     2010 &	31 & 57 & 208 & 157 \\
     2011 &	34 & 63 & 230 & 135 \\
     2012 &	35 & 65 & 238 & 127 \\
     2013 &	35 & 65 & 236 & 129 \\
     2014 &	30 & 65 & 238 & 127 \\
    \hline
     Average & 32 & 63 & 230 & 135 \\
    \hline
    \end{tabular}
\end{table}

\begin{figure*}[htbp]
\centering
\includegraphics[angle=0,scale=0.5]{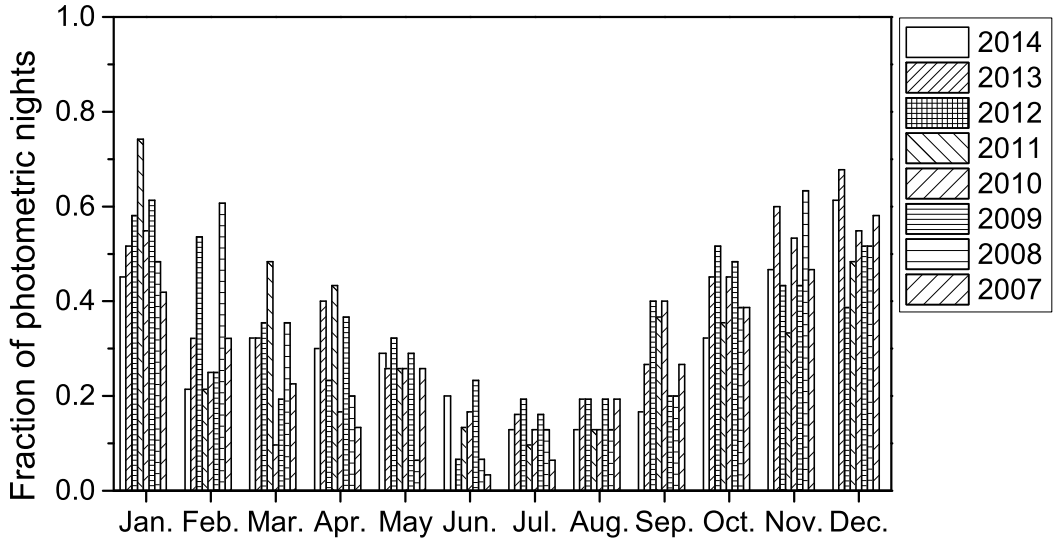}
\hfill \caption{Monthly distribution of photometric nights for the whole year during 2007-2014.}
\label{fig10}
\end{figure*}

\begin{figure*}[htbp]
\centering
\includegraphics[angle=0,scale=0.5]{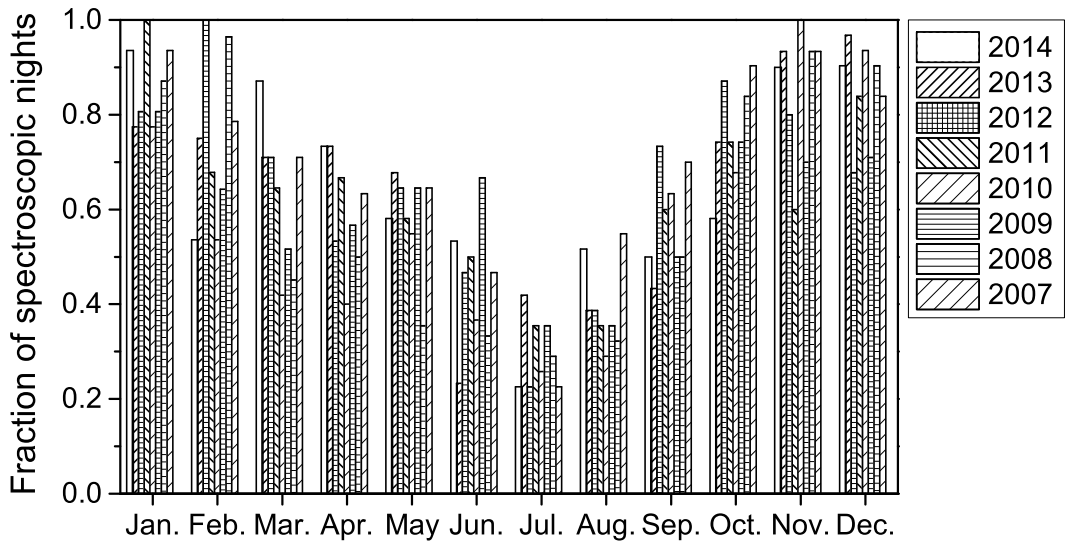}
\hfill \caption{Monthly distribution of spectroscopic nights for the whole year during 2007-2014.}
\label{fig11}
\end{figure*}

\begin{figure*}[htbp]
\centering
\includegraphics[angle=0,scale=0.5]{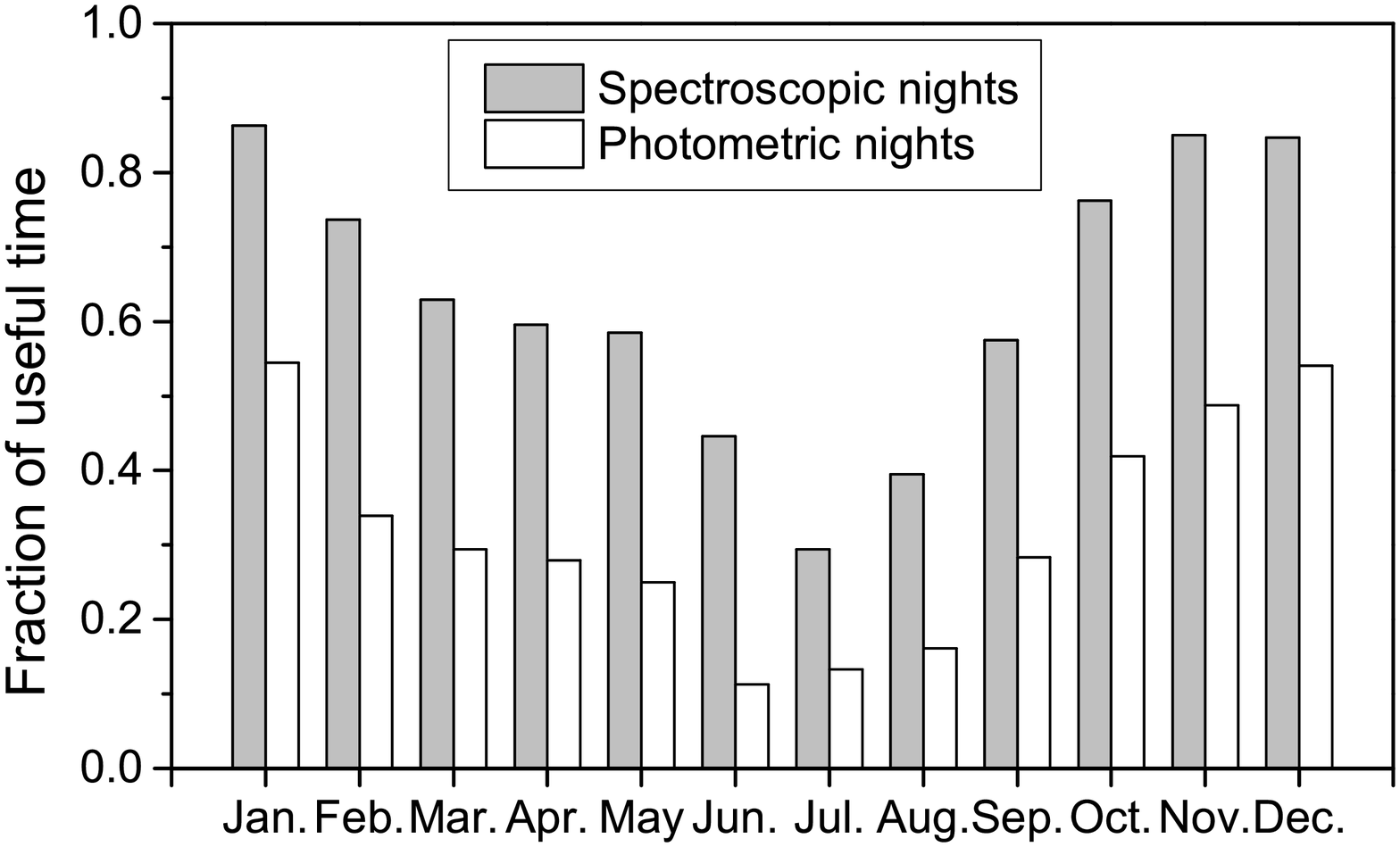}
\hfill \caption{Monthly average of photometric nights and spectroscopic nights for all years together from 2007 to 2014.}
\label{fig12}
\end{figure*}

In addition to the annual analysis, we have analysed monthly statistics of photometric nights and spectroscopic nights from 2007 to 2014 and showed in Figure~\ref{fig10} and Figure~\ref{fig11}, respectively. Seasonal distributions are obvious, autumn and winter are better in terms of fraction of photometric nights and spectroscopic nights. They have declining trend towards spring due to dust and wind, and goes to lowest in summer due to rains and high humidity. In order to analyze the relation between photometric nights and spectroscopic nights, we performed monthly average analysis of them from 2007 to 2014, and presented in Figure~\ref{fig12}. The fraction of useful time shows similar tendency and fraction of photometric nights decreases to the level of 10\% in summer.

\section{SEEING MEASUREMENTS FROM DIMM}

DIMM was first applied to measure seeing at Xinglong observatory in 2007 \citep{2010RAA....10.1061L}, twelve nights of DIMM data were collected for the seeing measurements at LAMOST site. They obtained the median seeing as 1.1$''$. Previous measurements of DIMM only have a short time span and could not reflect the monthly variations of seeing systematically.

DIMM seeing has been monitored from 2013 at Xinglong observatory systematically. The Dome for DIMM is on the 8th floor, just outside the LAMOST focal panel building. Its roof can slide to one side when it is operational. This dome design take fully into account the effect of wind, rain, snow and other environmental factors, and also taking into account the appearance of unity with LAMOST at the same time. This original DIMM was first deployed in 2011 using a portable 28~cm diameter Schmidt-Cassegrain telescope, with the focal-length of 2800 mm, equipped with a AVT Gruppy F-033 7.4 $\mu$m pitch 480$\times$640 pixel Sony ICX424 CCD camera. The altazimuth telescope was reformed to equatorial telescope for more stable tracking in the second year. The DIMM$'$s mask is made up of aluminum. There are two 50 mm apertures on the mask, one with prisms separated by 230 mm from the other sub-aperture. The control software and data reduction process are made to run on Ubuntu operating system. All softwares that are mentioned above are developed in C++ and Python language. Bright stars close to zenith are optimized observed objects. The image processing pipeline has fully considered the star's quality and position. CCD exposure time was set to 5~ms and each seeing value is measured by processing 800 images, and there are about 2 or 3 measurements for every minute.

\begin{figure*}[htbp]
\centering
\includegraphics[angle=0,scale=0.23]{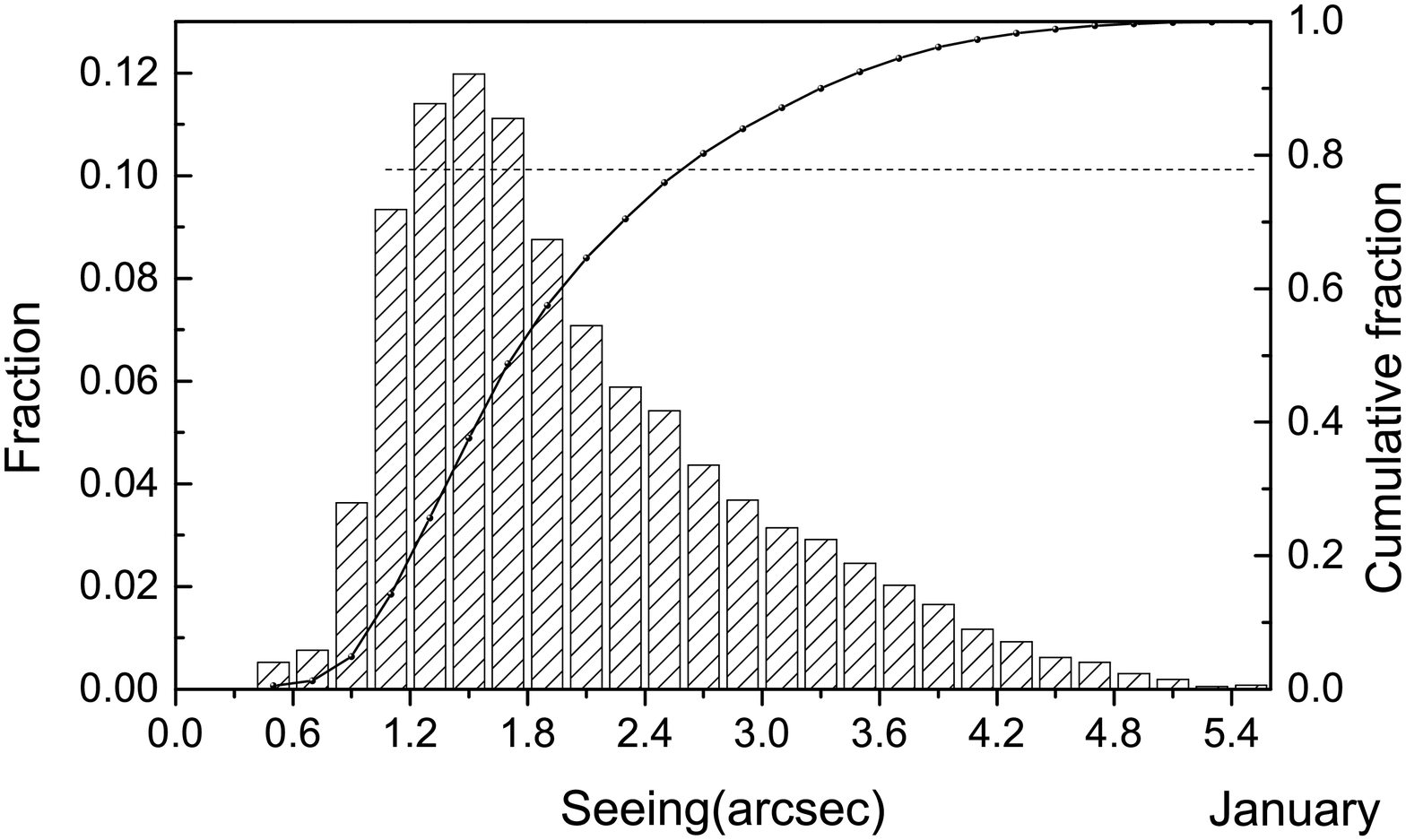}
\includegraphics[angle=0,scale=0.23]{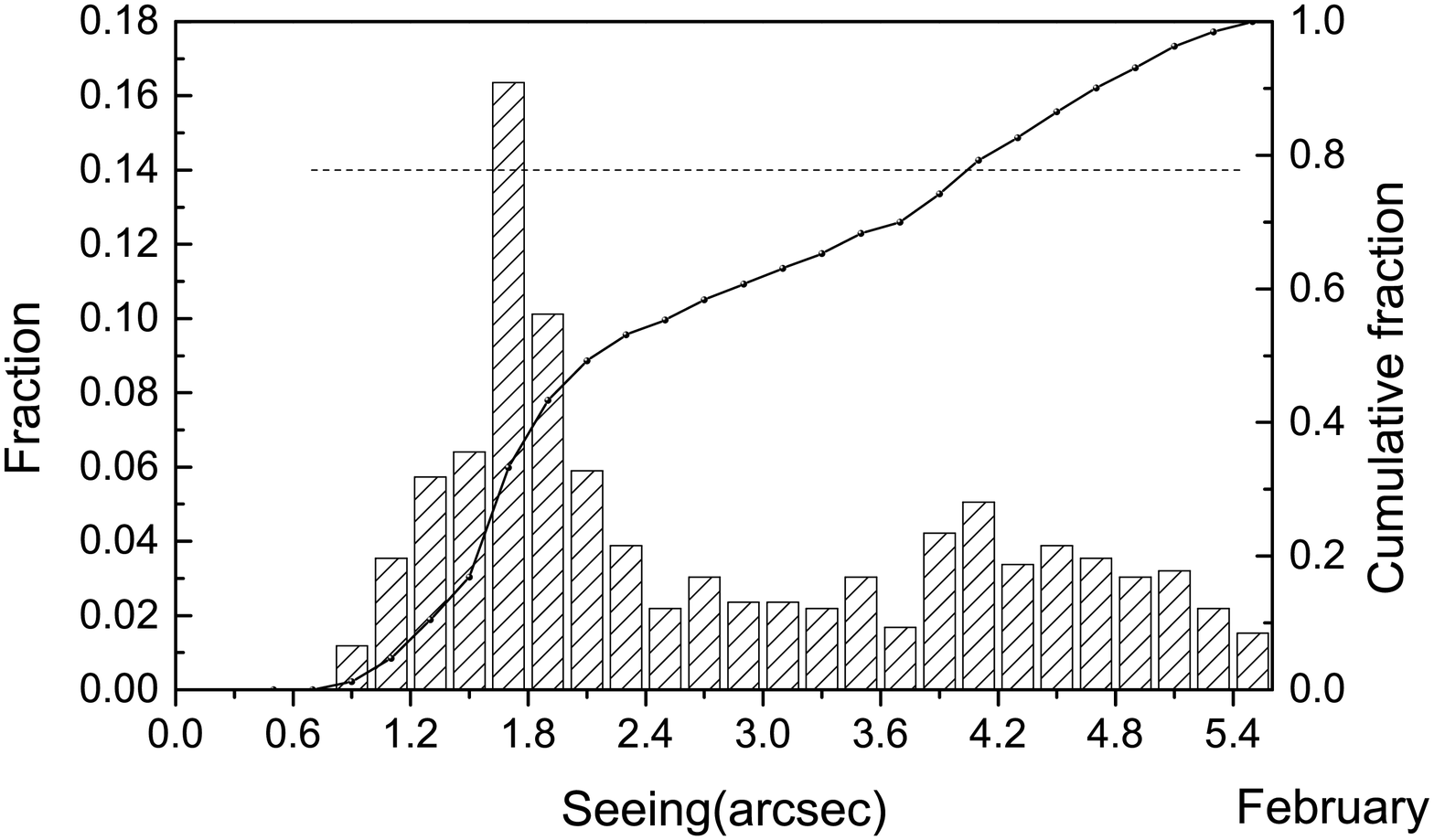}

\includegraphics[angle=0,scale=0.23]{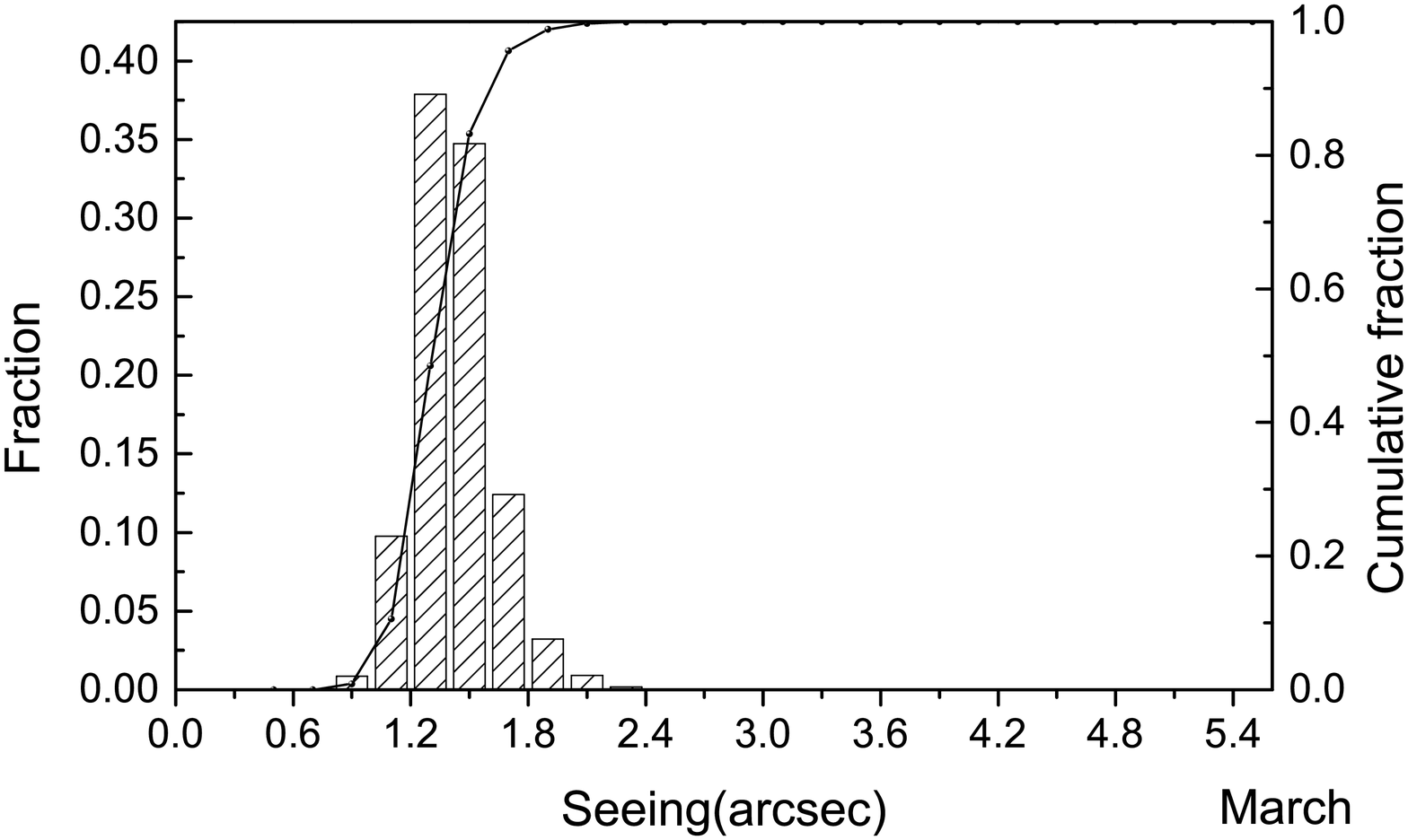}
\includegraphics[angle=0,scale=0.23]{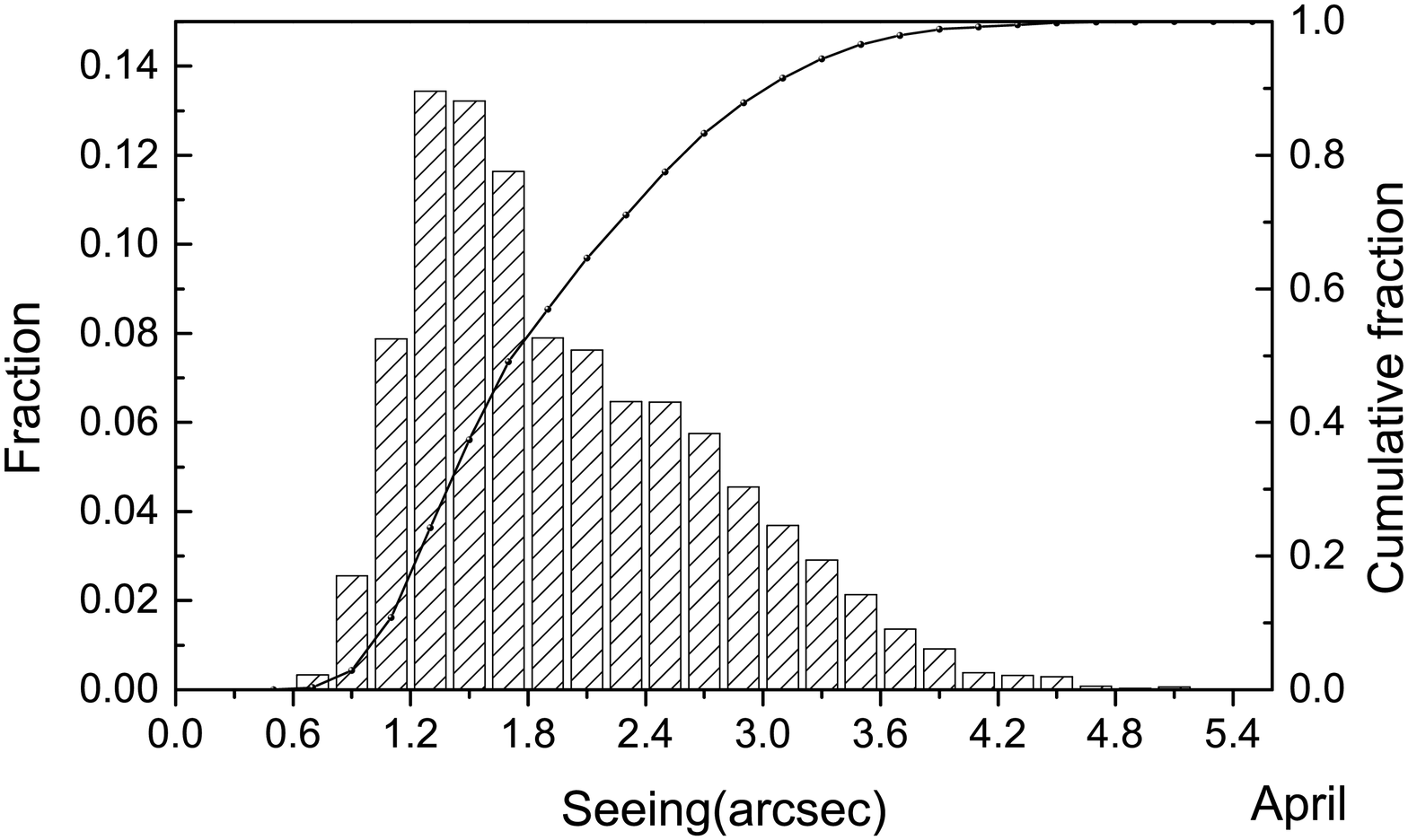}

\includegraphics[angle=0,scale=0.23]{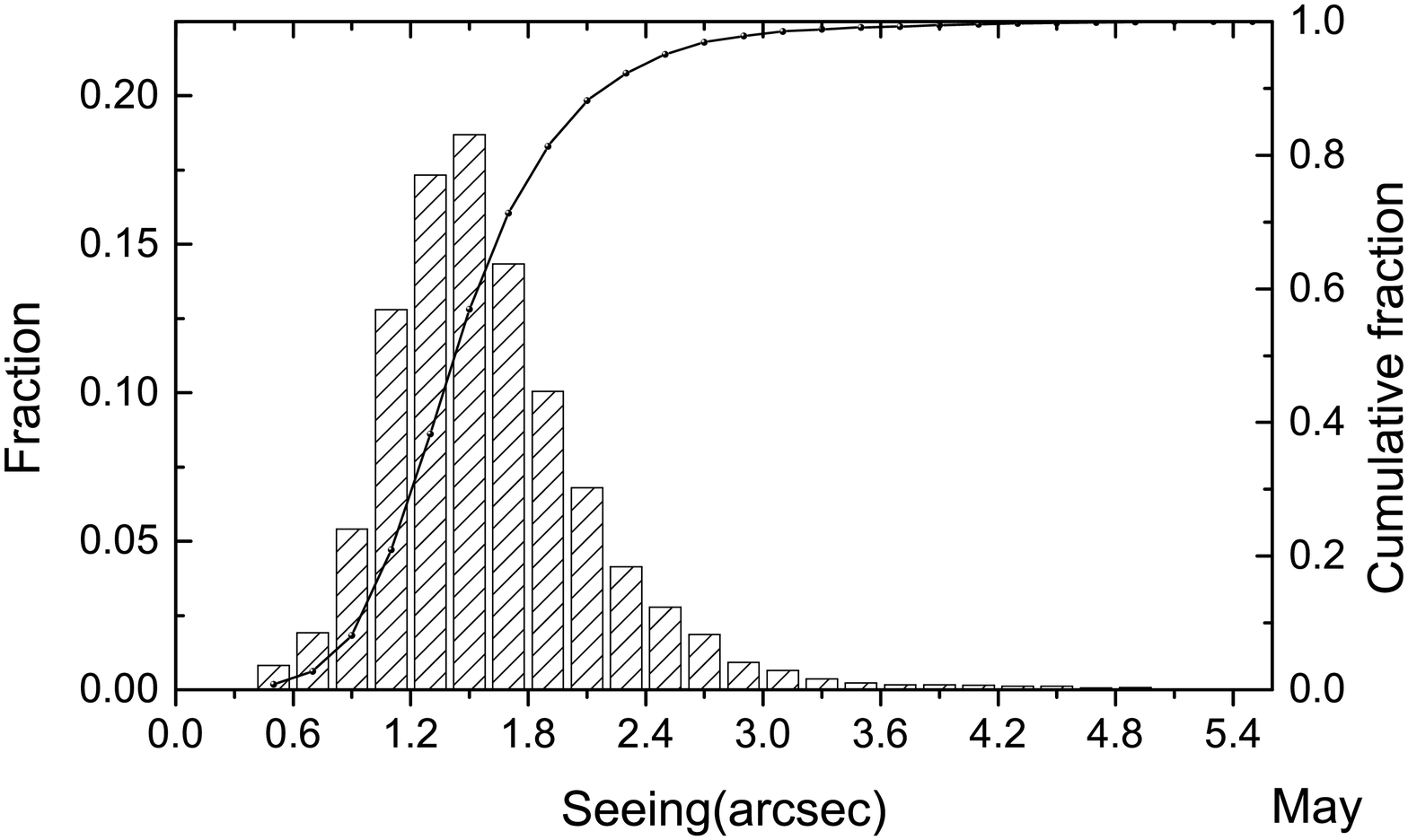}
\includegraphics[angle=0,scale=0.23]{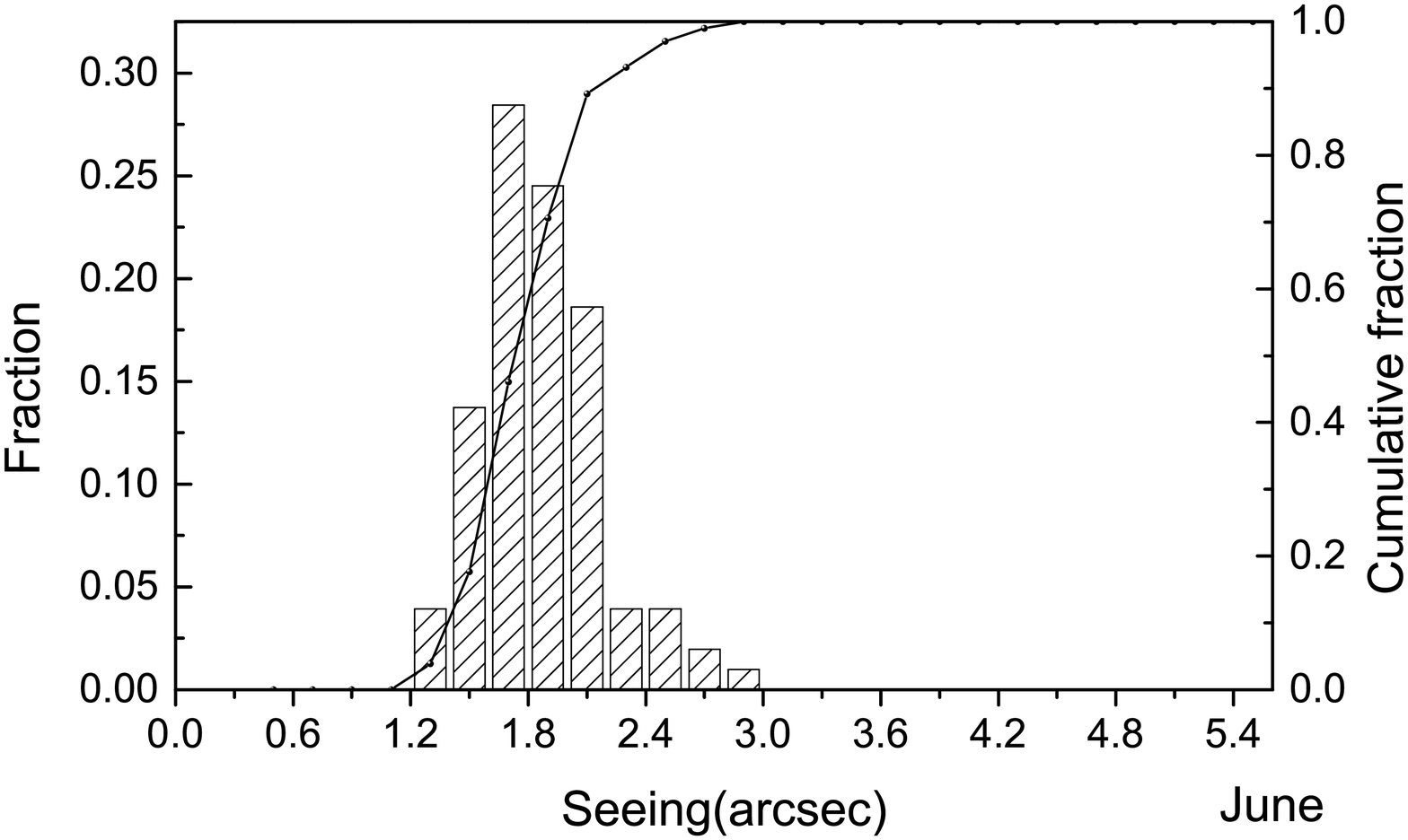}

\includegraphics[angle=0,scale=0.23]{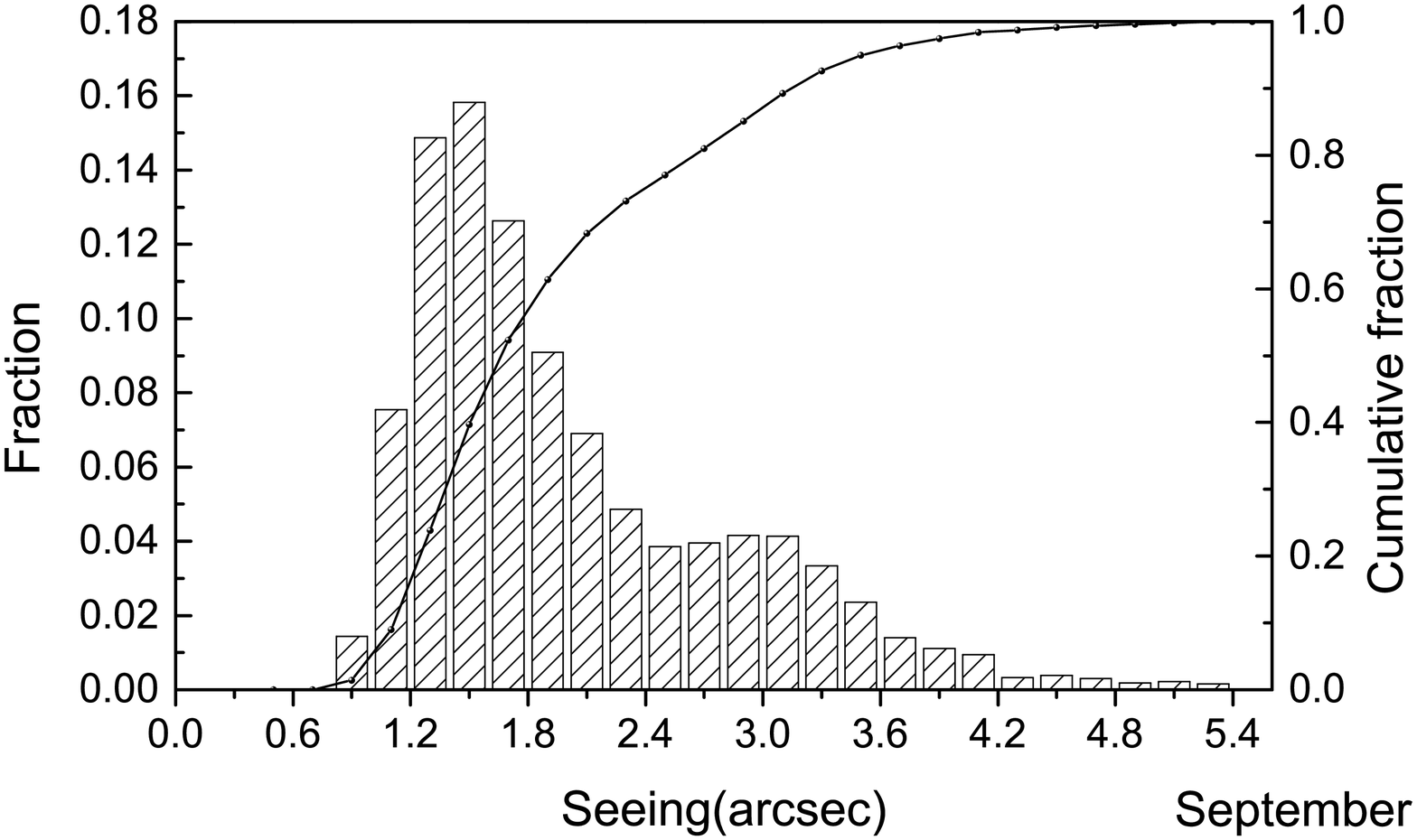}
\includegraphics[angle=0,scale=0.23]{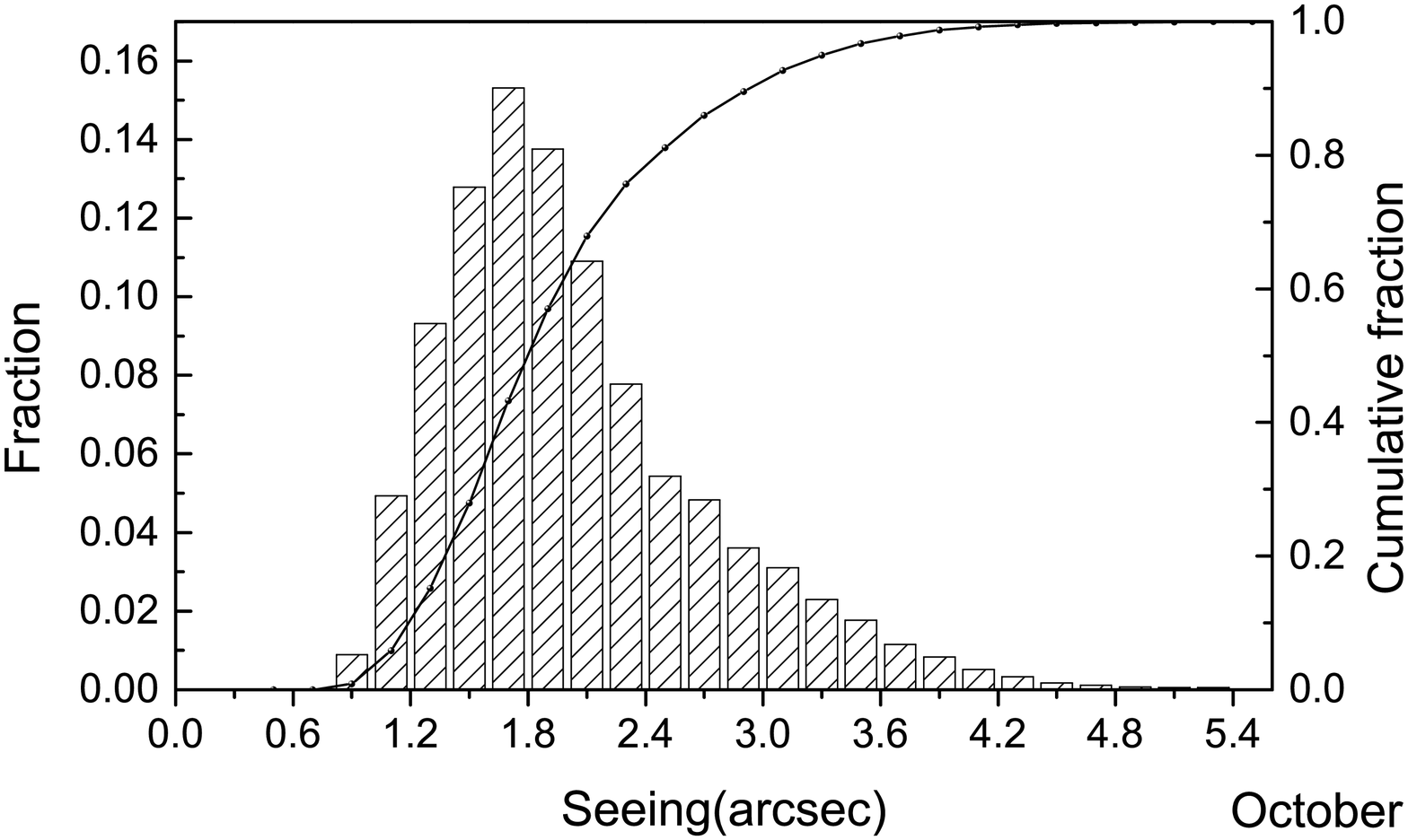}

\includegraphics[angle=0,scale=0.23]{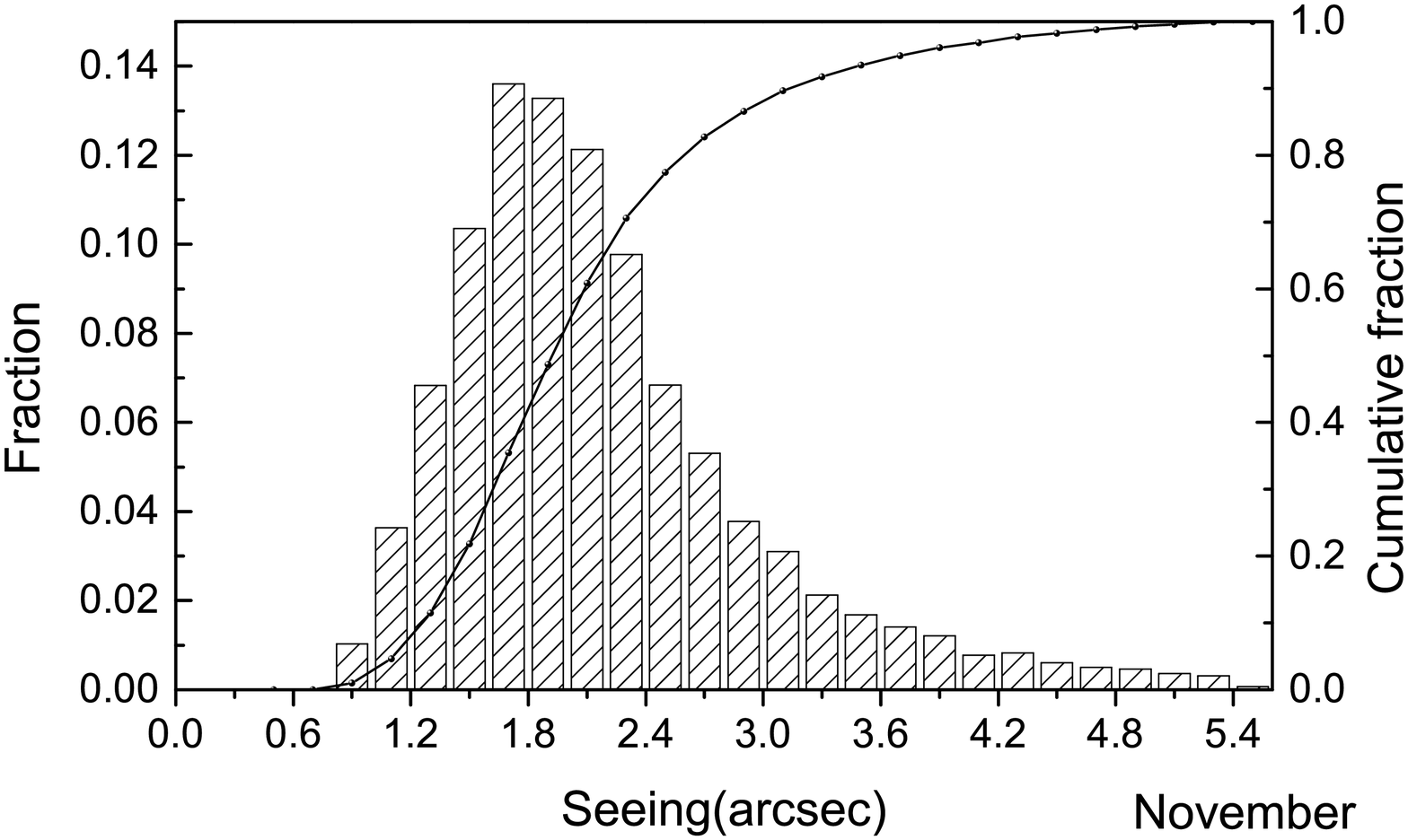}
\includegraphics[angle=0,scale=0.23]{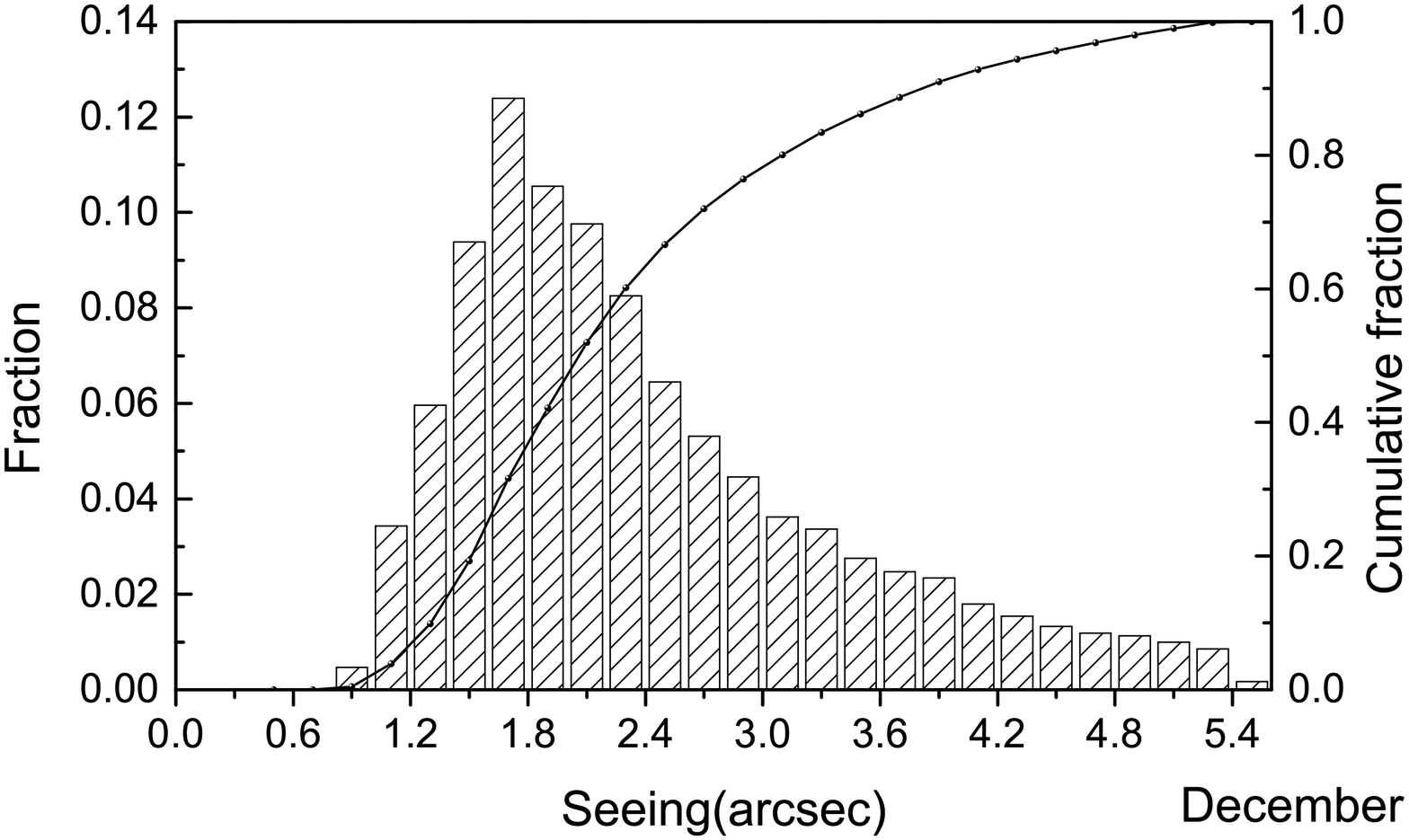}

\hfill \caption{Monthly cumulative distribution of seeing measurements from DIMM during 2014 are shown. Note that the data for July and August are missing due to telescope maintenance. Fractional and cumulative fractional values are mentioned in left and right of y-axis, respectively.}
\label{fig13}
\end{figure*}

We have obtained DIMM seeing data for whole year of 2014 and analysed on monthly basis. The cumulative distribution of each month, except July and August, are presented in Figure~\ref{fig13}. As the DIMM is installed on the LAMOST building, July and August are the maintained months for LAMOST, no data available in these two months. From the Figure we noticed the double peak distribution in February and September, which may be due to unstable winds.

\begin{figure*}[htbp]
\centering
\includegraphics[angle=0,scale=0.5]{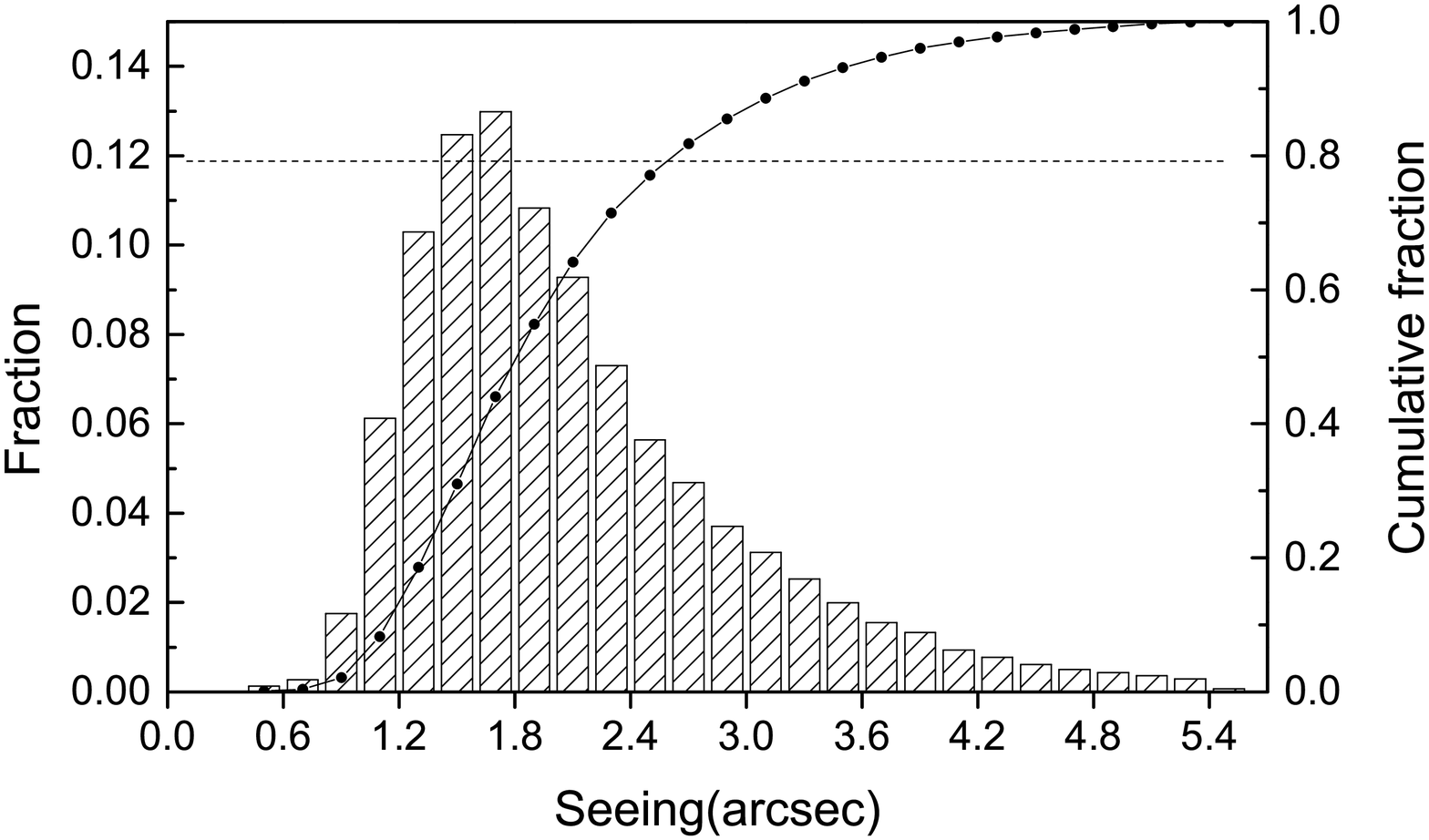}
\hfill \caption{Annual cumulative distribution of seeing for 2014.}
\label{fig14}
\end{figure*}

\begin{figure*}[htbp]
\centering
\includegraphics[angle=0,scale=0.5]{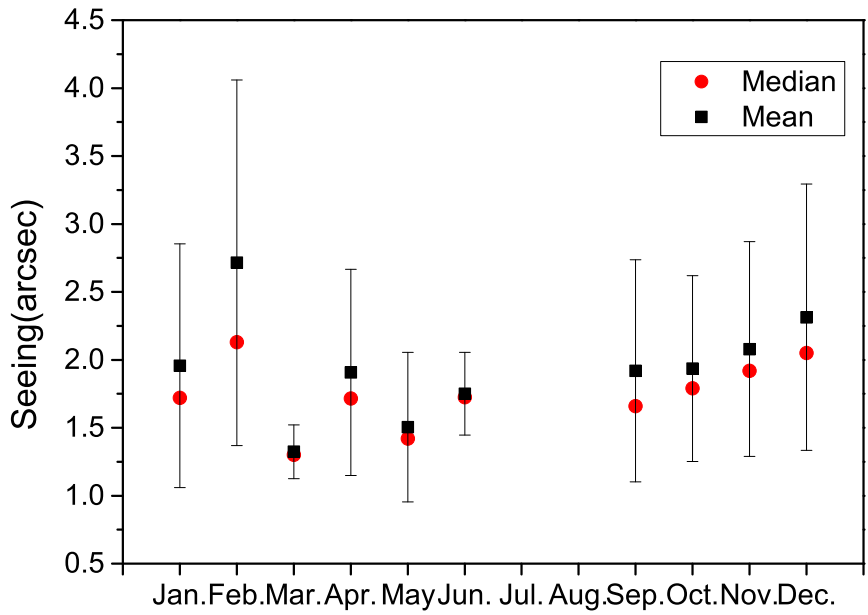}
\hfill \caption{Monthly mean and median of seeing values for 2014 are shown in black squares and red circles, respectively. Error bars indicates their standard deviation. See the electronic edition of the PASP for the color version of this figure.}
\label{fig15}
\end{figure*}

In addition, we present annual seeing distribution in 2014. Figure~\ref{fig14} shows 80\% of nights with seeing values are below 2.6$''$ whereas the distribution peaks around 1.8$''$. Mean and median seeing are around 1.9$''$ and 1.7$''$, respectively. Figure~\ref{fig15} presents the median, mean and its standard deviation of the monthly seeing value, which shows that the seeing distribution has the seasonal tendency, mean DIMM value is better in summer than winter, similar to previous studies on seeing distribution \citep{2012RAA....12..772Y}. Table 5 summarizes the data points used in analysis, median, mean and its standard deviation of monthly seeing data. We found that number of data points during February, March and June are very few, which maybe effected the shape of data distribution.

\begin{table}[htbp]

\centering
\caption{Monthly statistics of DIMM seeing measurements during 2014}
%\footnotesize
\begin{tabular}{ c c c c c }
	  \hline
  Month &Ndata  & Median  &  Mean   & Std\\
  &&($''$) & ($''$) & ($''$) \\
      \hline
January     & 17854  & 1.72   &  1.96 &   0.90 \\
February    & 593    & 2.13   &  2.71 &   1.34 \\
March       & 1650   & 1.30   &  1.32 &   0.20 \\
April       & 6242   & 1.72   &  1.91 &   0.76 \\
May         & 8718   & 1.42   &  1.50 &   0.55 \\
June        & 102	 & 1.73   &  1.75 &   0.30 \\
September   & 11745  & 1.66   &  1.92 &   0.82 \\
October     & 15521  & 1.79   &  1.94 &   0.68 \\
November    & 33214  & 1.92   &  2.08 &   0.79 \\
December    & 25068  & 2.05   &  2.31 &   0.98 \\
     \hline
Total & 120707 & 1.74  &  1.94  &    0.73 \\
     \hline
\end{tabular}
\end{table}

\section{THE SKY BRIGHTNESS OF XINGLONG OBSERVATORY}

The sky brightness of Xinglong observatory during 1995-2001 was studied by \cite{2003PASP..115..495L} and showed that the sky brightness of the North Pole field can reach up to 21.0 mag arcsec$^{-2}$. \cite{2012RAA....12..772Y} followed the same procedure and found that the sky brightness is consistent with 21.0 mag arcsec$^{-2}$. \cite{2012RAA....12.1585H} gave the sky brightness of Xinglong observatory based on the calibration observations of Tsinghua-NAOC Telescope (TNT), which showed that the sky brightness (in V-band) changed from 21.4 mag arcsec$^{-2}$ in 2005 to 20.2 mag arcsec$^{-2}$ in 2011. However, these data analysis did not consider the influence caused by different altitude and azimuth, as Xinglong observatory is surrounded by a number of nearby cities with light pollution and artificial acts. Study of the distribution with sky brightness and its variation on different directions due to light pollution plays a vital role for photometric telescopes at Xinglong observatory.

In this section we present the sky brightness of Xinglong observatory from Sky Quality Meter (SQM) and standard photometric measurements. The SQM is developed by the Unihedron company ($http://www.unihedron.com/projects/darksky/index.php$), which can measure the sky brightness in magnitudes per square arcsecond. This instrument has been widely used to measure the sky brightness for the GLOBE at Night program($http://www.globeatnight.org/sqm.php$). The SQM is installed at Xinglong observatory, fixed on the roof of an building which has no surrounding shade, pointing to zenith. Detail description of this SQM can be found from \cite{2013RAA....13.1255Y}.

\begin{figure*}[htbp]
\centering
\includegraphics[angle=0,scale=0.35]{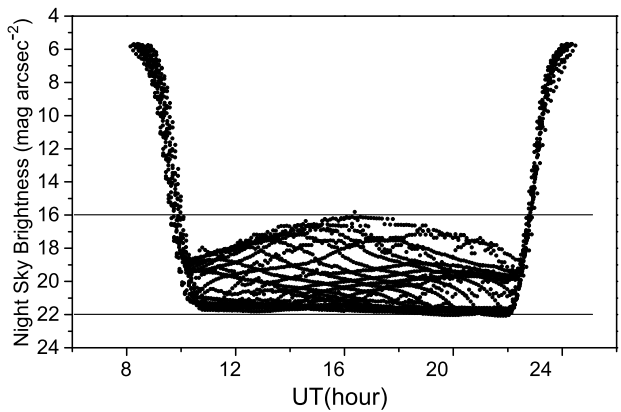}
\hfill \caption{Example data of SQM for one month are shown. Solid line at 22.0 mag arcsec$^{-2}$ and 16.0 mag arcsec$^{-2}$ represents moonless and moonlit night, respectively.}
\label{fig16}
\end{figure*}

SQM data are sampled every five minutes, in order to analyze the sky brightness variations with time and phase of moon, we collect one month data as an example, details are plotted in Figure~\ref{fig16}. It shows that the sky brightness gets darker after midnight, mainly due to the influence of city lights and the artificial acts. The sky brightness is also influenced by the moonlight, so we need to keep a large angular distance with the moon, in order to have a good SNR on faint target observations.

With the development of surrounding urban regions, light pollution at Xinglong observatory is getting more and more serious. Although sky brightness of the Xinglong observatory have been studied before, they only gave the brightness of the North Pole field or random area. In order to obtain the quantitative evaluation of the light pollution with different altitude and azimuth, we collect the data from TNT (an 0.8-meter Cassegrain reflecting telescope) by pointing it to the certain position on photometric nights. These data included photometric standard stars \citep{1992AJ....104..340L} with different airmass.

As CCD of TNT is cooled by Liquid nitrogen, the dark current is about 0.00025 e$^{-}$s$^{-}$pixel$^{-}$, which is negligible \citep{2012RAA....12.1585H}, so we have only corrected the bias and flat field for all object images. The entire data was reduced by following the standard procedure using various tasks in IRAF\footnote[1]{IRAF is distributed by National Optical Astronomy Observatory, which is operated by the Association of Universities for Research in Astronomy, Inc.(AURA) under cooperative agreement with the National Science Foundation.} and the Source Extractor software \citep{1996A&AS..117..393B}.

\begin{equation}
 M-m = ZP - \kappa  \chi
\end{equation}

Where $M$ is the catalogue magnitude, $m$ is the corresponding instrumental magnitude, which is calculated by the formula of $m = -2.5\;log_{10}(\frac{counts}{t_{exp}})$, $\kappa$ is the extinction coefficient, $\chi$ is the airmass of image. We have followed the similar photometric procedure by \cite{2014PASP..126..496G}.

\begin{figure*}[htbp]
\centering
\includegraphics[angle=0,scale=0.35]{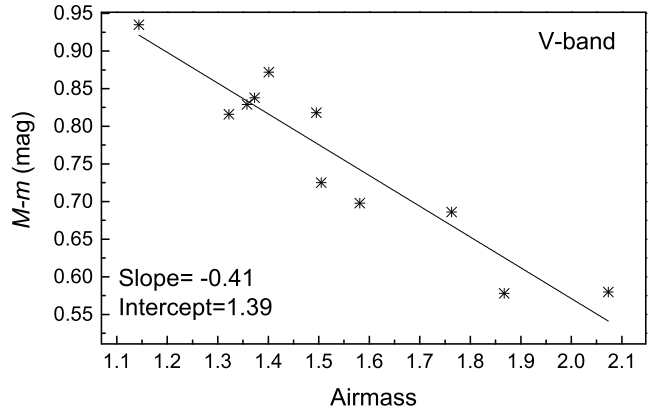}
\hfill \caption{Example of the linear regression on data during photometric night, in order to obtain zero-point on the V-band.}
\label{fig17}
\end{figure*}

After the photometric calibration, we got the zero-point ($ZP$) of the instrument from the linear regression shown in Figure~\ref{fig17}. The pixel scale of the detector is 0.516$\arcsec$ pixel$^{-1}$. We have calculated the sky brightness using
\begin{equation}
%\label{eq:skyb}
M_{NSB} =ZP-2.5\;log(\frac{sky_{count}}{scale^2 \; t_{exp}})
\end{equation}

Where $ZP$ is obtained from the photometric calibration mentioned above. $sky_{count}$ is the sky flux per pixel, we removed the influence of stars and cosmic rays and gave the median value of background as the $sky_{count}$. $scale$ is calculated with the focal length of TNT and the pixel size of detector, e.g. 0.516$\arcsec$ pixel$^{-1}$, $t_{exp}$ is the exposure time (in seconds).
\begin{figure*}[htbp]
\centering
\includegraphics[angle=0,scale=0.4]{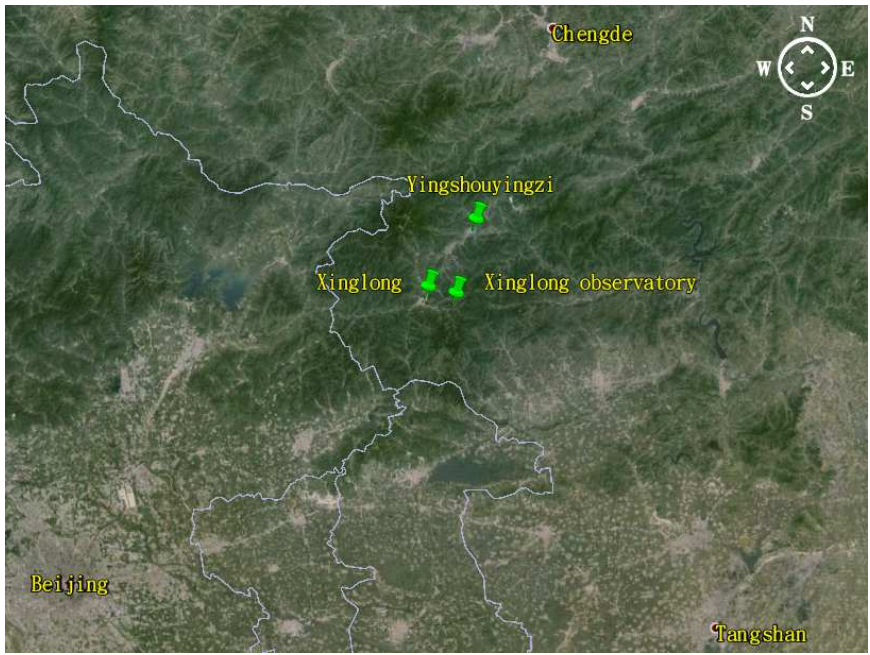}
\hfill \caption{Cities and counties causing light pollution around Xinglong observatory are marked in the map. Beijing in southwest, Tangshan in southeast, Xinglong in west, Chengde and Yingshouyingzi in northeast. See the electronic edition of the PASP for a color version of this figure.}
\label{fig18}
\end{figure*}

\begin{figure*}[htbp]
\centering
\includegraphics[angle=0,scale=0.4]{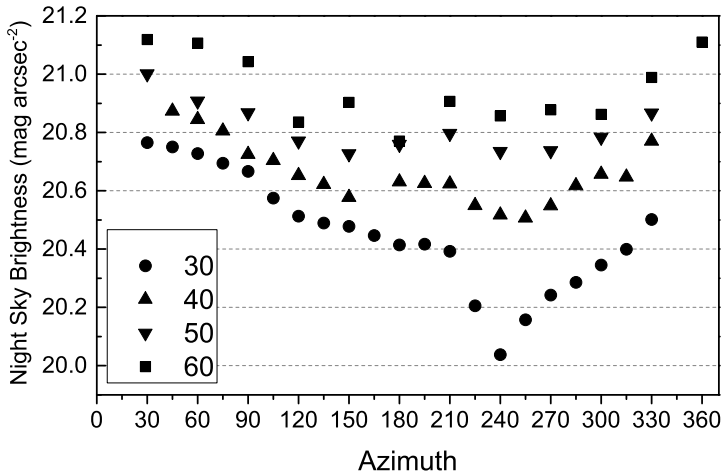}
\hfill \caption{Distribution of sky brightness (V-band) with different azimuth and altitude at Xinglong observatory. $0^{\circ}$ corresponds to north and rest follows clockwise direction. Circles, triangles, inverted triangles, and squares indicate the altitude of $30^{\circ}$, $40^{\circ}$, $50^{\circ}$, $60^{\circ}$, respectively.}
\label{fig19}
\end{figure*}

We noticed that Xinglong observatory is mainly surrounded by three cities (e.g. Beijing, Chengde and Tangshan), and two counties (Xinglong and Yingshouyingzi) that are shown in Figure~\ref{fig18}. According to our analysis, counties are very near to observatory and have little influence compare to cities. Beijing is located at southwest direction of Xinglong observatory, which corresponds mainly with azimuth of $240^{\circ}$, Chengde is located at northeast direction of Xinglong observatory, Tangshan is located at southeast direction of Xinglong observatory. We test the influence of the light pollution with different altitude and azimuth, the detailed distribution of sky brightness is shown in Figure~\ref{fig19}. We find the sky brightness at the zenith is at the level of 21.1 mag arcsec$^{-2}$, which is comparable with the sky brightness about 21.0 mag arcsec$^{-2}$ (V-band) measured by \cite{2003PASP..115..495L} until 2001 and \cite{2012RAA....12..772Y} until 2011 at Xinglong observatory. Light pollution of Beijing is more prominent, especially at altitude of $30^{\circ}$ and $40^{\circ}$, which could reach to the level of 20.0 mag arcsec$^{-2}$ at the altitude of $30^{\circ}$. The light pollution of Tangshan is prominent at higher altitude, almost the same as Beijing. Our analysis of the influence of light pollution can be used as a reference to the observers when they use telescopes to observe faint objects at Xinglong observatory.

\section{CONCLUSIONS}

We have made an attempt to understand the astronomical observing conditions at Xinglong observatory, using the data for a period of 8 years from 2007 to 2014, by performing the analysis of three important parameters: meteorological information, seeing and sky brightness.

Annual statistics of air temperature show that it is almost constant during the period analysed and ranged from 9.2$^\circ$C to 33.7$^\circ$C. Monthly statistics show that air temperature in summer is higher compare to other seasons, which can reach around 30$^\circ$C in daytime and go down to 10$^\circ$C in the nighttime. In order to stabilize the temperature and airflow at the beginning of observations, the ventilation device and the dome slit are opened at least one hour before the observation everyday. We have also checked the variation of air temperature during overnight through the analysis of annual nighttime air temperature and found that the temperature gradient is almost stable, which indicates better dome seeing and the quality of the images.

Annual statistics of wind speed show that it is under 4~m s$^{-1}$ most of the time, the peak value of distribution is around 2~m s$^{-1}$ during these years. Monthly statistics suggested that mean and median value of wind speed are almost constant during the period analysed and ranged from 1.0 m s$^{-1}$ to 3.5 m s$^{-1}$. Analysis suggested that wind speed above 15 m s$^{-1}$, which is safety limit for operating telescopes at Xinglong observatory, happened to be more during winter and spring. Humid time occurs mostly in summer and it is around 55\%, due to the monsoon climate, which is main cause for lost time during summer. We found that it also extends a small fraction to autumn mainly during September.

Through the analysis of meteorological information at Xinglong observatory, we find that the fraction of photometric nights and spectroscopic nights are almost constant from 2007 to 2014, without obvious variation. Average percentage of spectroscopic nights is 63\% per year. Over 60\% of nights are suitable for observation. Unuseful time are mainly distributed in summer as it is the rainy season at Xinglong observatory. Cloud, high humidity, wind and dusts also influence observations. Average percentage of photometric nights is 32\% per year. Distribution of photometric nights and spectroscopic nights show similar seasonal trend. Fraction of photometric nights decreases to the level of 10\% in summer.

Seeing data was collected from DIMM. Annual statistics of seeing shows that median and mean seeing are around 1.7$''$ and 1.9$''$, respectively. Peak of seeing distribution is around 1.8$''$, 80\% of the seeing is better than 2.6$''$. Seeing is seasonal dependent, and better in summer than winter.

Sky brightness data were collected from SQM and photometric observation. Sky brightness gets darker after the midnight from the SQM, which means the city lights and artificial acts have influence on the light pollution. We find that the sky brightness is influenced by the moonlight, so astronomers need to keep a large angular distance with the moon when observing faint targets. Standard photometric measurement by TNT with different azimuth and altitude on moonless night shows that sky brightness at the zenith is at the level of 21.1mag arcsec$^{-2}$. However, it becomes brighter at large zenith angle due to the light pollution of surrounding cities, Light pollution of Beijing is prominent at the altitude of $30^{\circ}$ and $40^{\circ}$, which could reach to the level of 20.0 mag arcsec$^{-2}$ at altitude of $30^{\circ}$. Light pollution of Tangshan is prominent at higher altitude, almost the same as Beijing.

Our detail analysis on above parameters resulted the number of useful time and night sky brightness are almost consistent during the period, suggesting the site is stable and suitable for optical astronomical observations. Our analysis towards the astronomical observing conditions at Xinglong observatory can be used as a reference to the observers on targets selection, observing strategy, and telescope operation.

We are thankful to anonymous referee for his/her valuable comments and suggestions that lead to improve the manuscript. We thank the supports of the night assistants at Xinglong Observatory. We thank Bharat Kumar Yerra for his observant comments on the manuscript that lead to a better presentation. This work is partly supported by National Natural Science Foundation of China under grant No.11373003 and National Key Basic Research Program of China (973 Program) No. 2015CB857002.

\end{document}